\documentclass{desyproc}
\newcommand{\DS}[1]{/\!\!\!#1}
\newcommand{\be}{\begin{equation}}
\newcommand{\ee}{\end{equation}}
\newcommand{\ba}{\begin{eqnarray}}
\newcommand{\ea}{\end{eqnarray}}
     
\begin{document}

\title{\vspace{-1.8cm} 
\begin{flushright} {\small SI-HEP-2013-15}
\end{flushright}
Applications of QCD Sum Rules \\to Heavy Quark Physics}

\author{{\slshape Alexander Khodjamirian}\\[1ex]
Theoretische Physik 1, 
Naturwissenschaftlich-Technische Fakult\"at,\\
Universit\"at Siegen, D-57068 Siegen, Germany
\footnote{Lectures at the Helmholtz International Summer School ``Physics of Heavy Quarks and Hadrons'', \\ July 2013, Dubna, Russia}}

\contribID{xy}

\desyproc{DESY-PROC-2013-03}
\acronym{HQ2013} 

\maketitle

\begin{abstract}
In these lectures,  I  present several important applications 
of QCD sum rules to the decay processes involving 
heavy-flavour hadrons. The first lecture  is introductory.
As a study case, the sum rules 
for decay constants of the heavy-light mesons are considered. They are 
relevant for the leptonic decays of $B$-mesons.
In the second lecture I describe the method of QCD light-cone 
sum rules used to calculate the heavy-to-light form factors at large 
hadronic recoil, such as the $B\to \pi \ell \nu_\ell$ form factors. In the third 
lecture,  the nonlocal hadronic  amplitudes in the flavour-changing neutral current decays
$B\to K^{(*)}\ell\ell$ are discussed.  Light-cone sum rules provide important 
nonfactorizable contributions to these amplitudes.
\end{abstract}

\section*{Introduction}

The method of sum rules in quantum chromodynamics (QCD) 
developed in \cite{SVZ}  relates hadronic parameters, such as decay constants
or transition form factors, 
with the correlation functions of quark currents.
Let me outline the three key elements of this method:
\begin{itemize}

\item {\bf Correlation function of  local quark currents} is defined.
The simplest, two-point correlation function is formed by two quark-antiquark current
operators sandwiched between the QCD vacuum states. This is a function of 
the 4-momentum transfer between 
the currents.  In the region of 
large spacelike momentum transfers, 
the correlation function 
represents a short-distance fluctuation of quark-antiquark fields. 
The  propagation of quarks and antiquarks 
at short distances is asymptotically  
free, the gluon exchanges being suppressed by a 
small QCD coupling. In addition,
the interactions  with ``soft'' (low momentum) 
quark-antiquark and gluon fields populating  the QCD vacuum
have to be taken into account. 

\item {\bf Operator-product expansion (OPE)} of the correlation function 
is worked out. This expansion provides an analytical
expression for  the correlation function at spacelike momentum transfers, 
with a systematic separation 
of short- and long-distance effects. The former are described by 
Feynman diagrams with quark
and gluon propagators and vertices, whereas the latter 
are encoded by universal parameters related to the nonperturbative 
QCD dynamics. In the case of two-point sum rules, these parameters are the averaged local densities of the QCD vacuum fields, the condensates. 
The contributions of vacuum effects in OPE are suppressed by inverse powers 
of the large momentum and/or heavy-quark mass scale, allowing one to truncate the expansion  at some maximal power. 

\item{\bf Hadronic dispersion relation} for the correlation function
is employed. 
The basic unitarity condition allows one to express the 
imaginary part (spectral density)  of the correlation function in terms 
of the sum and/or integral over all intermediate 
hadronic states with the quantum numbers of the quark currents.
On the other hand, employing the analyticity of the correlation 
function in the momentum transfer variable,  one relates 
the OPE result at spacelike momentum transfers 
to  the integral over hadronic 
spectral density. In this way a  link between QCD and hadrons
is established, and the resulting relation between the OPE expression 
and hadronic sum  is naturally called a ``QCD sum rule''.
 \end{itemize}

After this general description of the method, let me quote a shorter but more  
emotional definition of QCD sum rules:
"Snapshots of hadrons or the story of how the vacuum medium determines the 
properties of the classical mesons which are produced, live and die in the QCD 
vacuum", given as a title to the review \cite{Shifman98}
written by one of the founders of this method.

Due to a vast amount of applications of QCD sum rules accumulated during 
many years, these lectures represent
only a brief guide to the field,
exemplifying applications 
to a few important processes involving heavy flavoured hadrons.  
More detailed  reviews  are listed in \cite{Shifman98,reviews,KR98,CK,Braun_rev97}.

\section{\!\!\!\!\!\!.~Lecture: Calculating the $B$-meson decay constant}

In this introductory lecture, I consider, as a study case, the 
QCD sum rule derivation for an important
hadronic parameter -- the $B$-meson decay constant.

\subsection{ $B$-meson leptonic decays}

\begin{figure}[hb]
\centerline{
\includegraphics[width=0.35\textwidth]{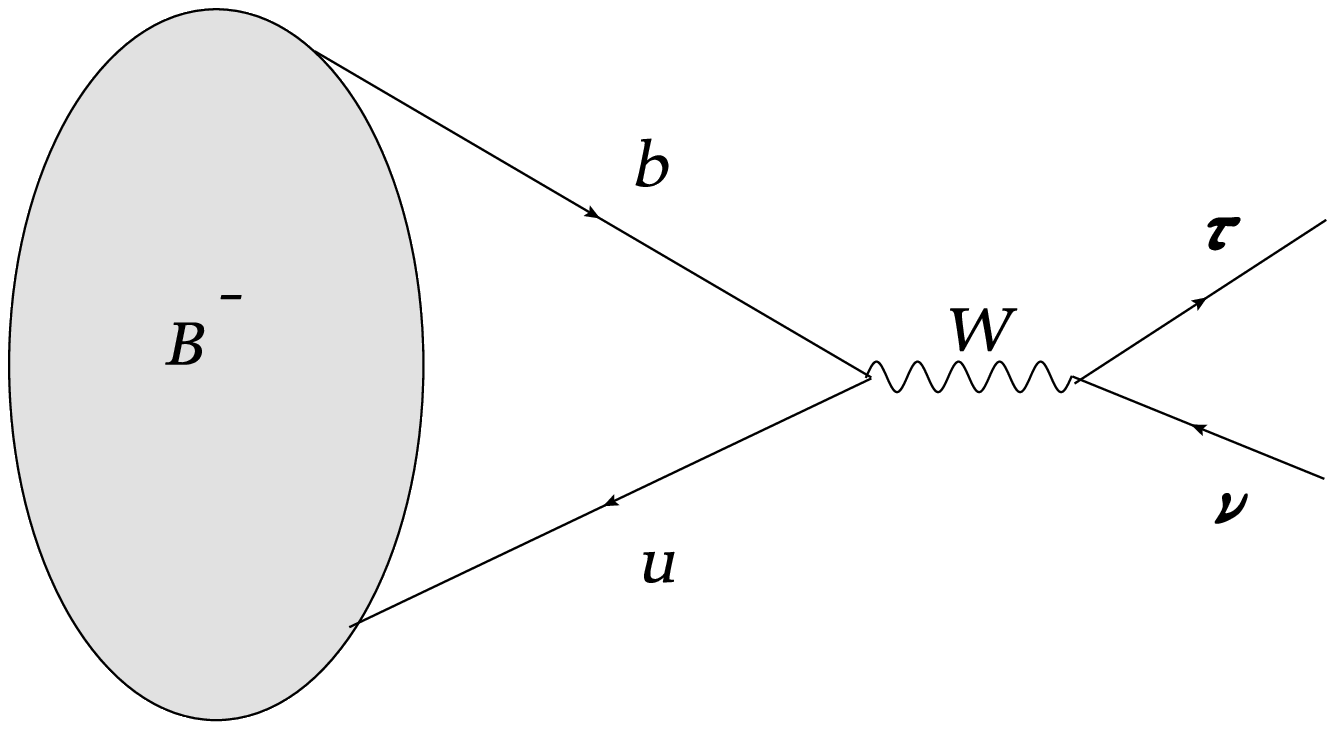}\hspace{1cm}
\includegraphics[width=0.35\textwidth]{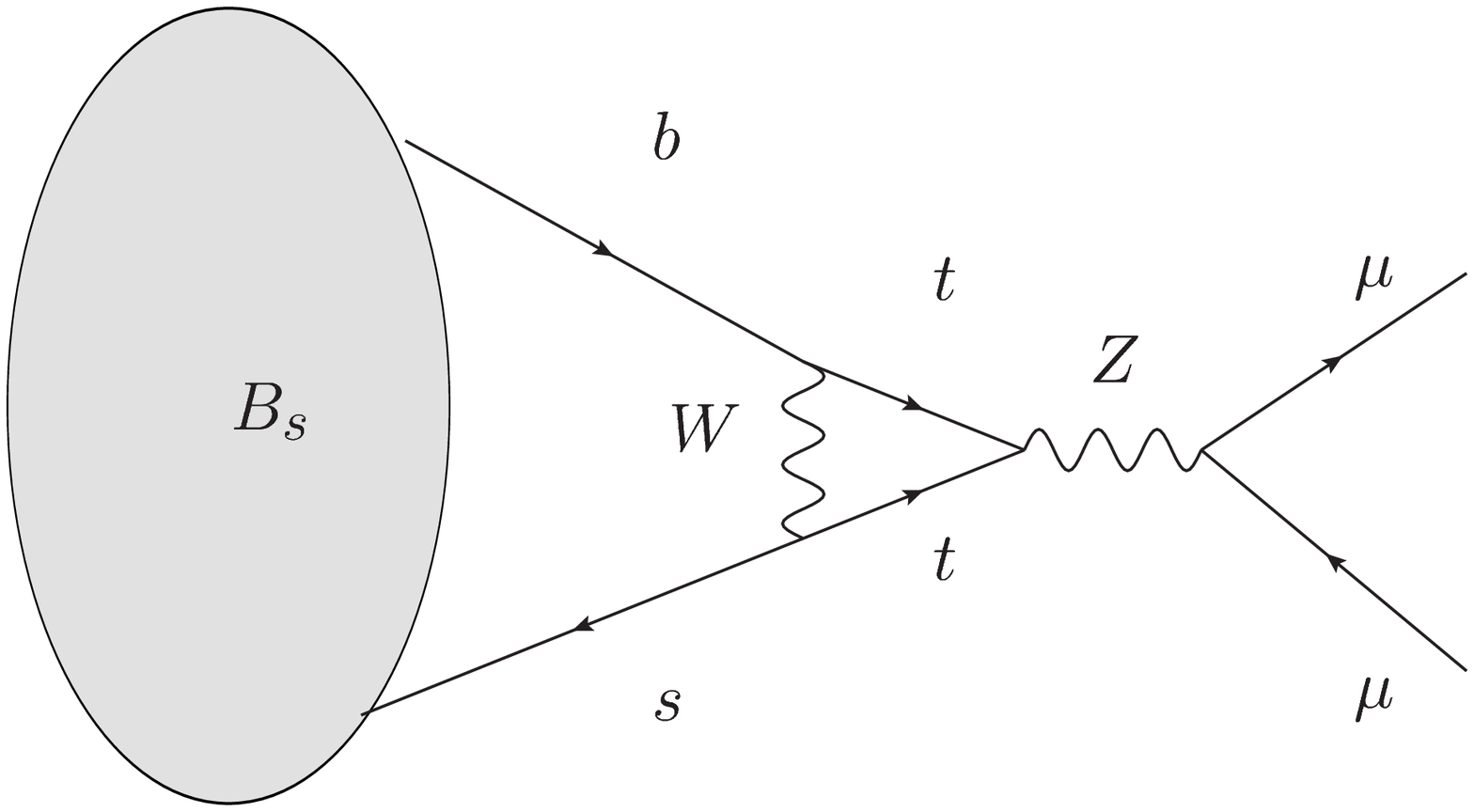}
}
\begin{center}
{\small (a) \hspace{5.4cm} (b)}
\end{center}
\caption{\small (a)  Diagram of the weak  leptonic decay
$B^-\to\tau\bar{\nu}_\tau$;  (b)  one of the diagrams of 
the FCNC leptonic decay $ B_s\to \mu^+\mu^-$. The initial $B$ meson
is denoted by a blob. }
\label{fig:Bleptonic}
\end{figure}

The decay diagrams are shown in Fig.~\ref{fig:Bleptonic}. The first 
leptonic decay is a weak transition  $B^-\to \ell\bar{\nu_\ell}$ via virtual $W$ boson exchange. 
For $\ell=\tau$ its branching fraction was measured at $B$ factories \cite{BelleBtaunu}. 
The second decay, $\bar{B}_s\to \ell^+\ell^-$, is a rare 
flavour-changing neutral current (FCNC) transition  
generated by the loop diagrams with heavy particles ($t$,$Z$,$W$). 
Its recent observation at LHC \cite{LHCbBsmumu} was a great experimental achievement.
Although short-distance electroweak interactions 
are quite different, these decays  have one common feature:
the initial $B$-meson annihilates and the final state contains no hadrons
i.e. it is a vacuum (lowest energy) state of QCD.
The decay amplitude of the weak decay in Standard Model (SM):
\begin{equation}
A(B^-\to \tau^- \bar{\nu}_\tau)=\frac{G_F}{\sqrt{2}}\,V_{ub}\,
\bar{\tau}\gamma^\mu(1-\gamma_5) \nu_{\tau}
\langle 0|\bar{u}\gamma_\mu\gamma_5 b| B^- \rangle\, ,
\label{eq:Bleptampl}
\end{equation}
contains the simplest possible hadronic matrix element 
\begin{equation}
\langle 0|\bar{u}\gamma^\mu\gamma_5 b| B (p_B)\rangle=i p^\mu_B  f_B,   
\label{eq:fBdef}
\end{equation}
in which the local operator of $b\to u$ weak transition current 
is sandwiched between $B$ and the vacuum state.
The above formula in terms of  a constant parameter $f_B$ reflects the fact 
that $p_B^\mu$ is the only 4-momentum involved in this
hadronic matrix element  and $p_B^2=m_B^2$. The quantity $f_B$ is the 
$B$-meson decay constant we are interested in. 
In order 
to use  the experimental measurement of the decay branching fraction:   
\begin{equation}
BR(B^-\!\to \!\tau^- \bar{\nu}_\tau)= 
\frac{G_F^2 |V_{ub}|^2}{8\pi}m_\tau^2m_B 
\left(1-\frac{m_\tau^2}{m_B^2}\right)^2
f_B^2\tau_{B^-}\,,
\label{eq:BrBtaunu}
\end{equation}
where $\tau_{B^-}$ is the lifetime of $B^-$, 
one  needs to know $f_B$ from the theory.
This will allow  one to 
to extract the fundamental CKM parameter $|V_{ub}|$ 
or to check if there is an admixture of new physics, e.g., of a charged 
Higgs boson exchange, in this decay. 

The rare leptonic decay, $B_{s}\to \mu^+\mu^-$ , 
is  even more  sensitive to new physics  contributions,
due to the presence  of heavy  particle loops. 
The corresponding hadronic matrix element  
\begin{equation}
\langle 0|\bar{s}\gamma^\mu\gamma_5 b| B_s (p_B)\rangle=i p^\mu_Bf_{B_s}
\label{eq:fBs}
\end{equation}
is very similar to Eq.~(\ref{eq:fBdef}), and the squared decay constant
$f_{B_s}^2$ enters the decay width.
The CKM suppressed $B_d\to \mu^+\mu^-$ decay contains
the decay constant of $B_d$. Due to isospin symmetry between $u$ and $d$ quarks,
$f_{B_d}\simeq f_{B_u} \equiv f_{B}$ with a good accuracy. On the other hand,
$f_{B_s}$ and $f_B$ noticeably differ, 
because the $SU(3)_{flavor}$ symmetry is violated by the quark mass difference   $m_s-m_{u,d}$.
Hence, an accurate calculation of $f_{B_s}$ has to take into account the finite
$s$-quark mass.

We conclude that a QCD calculation of $f_B$   
is indispensable for disentangling the fundamental  
flavour-changing transitions from the measurements  of leptonic $B$ decays.

\subsection{ $B$-meson  decay constant in QCD }
\begin{wrapfigure}{r}{0.25\textwidth}
\centerline{
\includegraphics[width=0.18\textwidth]{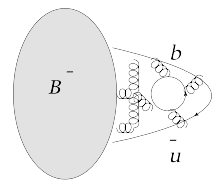}\hspace{1cm}
}
\vspace{-0.1cm}
\caption{\small $B$ meson transition to vacuum.}
\label{fig:Bann}
\end{wrapfigure}

The task is to calculate the hadronic matrix element 
(\ref{eq:fBdef}) which is shown in Fig.~\ref{fig:Bann},
separated from the electroweak part of the leptonic decay amplitude.
The wavy lines and loops in this figure indicate 
gluons and quark-antiquark pairs interacting with the valence
$b$ and $\bar{u}$ quarks inside $B^-$ meson. But 
these lines and loops are only illustrative:  it is not possible to directly attribute 
QCD Feynman graphs to a hadronic amplitude.

\begin{wrapfigure}{r}{0.65\textwidth}
\centerline{
\includegraphics[width=0.45\textwidth]{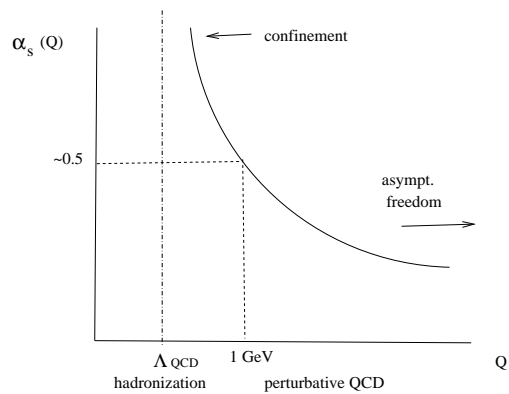}\hspace{1cm}
}
\caption{\small Dependence  of the effective coupling in QCD on the energy/momentum
scale $Q$.}
\label{fig:alphas}
\end{wrapfigure}

The quantum field theory 
of quarks, gluons and their interactions
is encoded in the QCD Lagrangian: 
\begin{eqnarray}
L_{QCD}(x)=-\frac14G_{\mu\nu}^{{a}}G^{{a} \,\mu\nu}(x)
+ \sum_{q=u,d,s,c,b,t}\!\!\!\!\!\bar{q}^{\, i}(x)(iD_\mu\gamma^\mu-m_q)q^{\,i}(x)
\label{eq:Lqcd}
\end{eqnarray}
where $D_\mu=\partial_\mu-ig_s\frac{\lambda^a}2A^a_\mu $  is the covariant derivative, 
$G_{\mu\nu}^{{a}}=\partial_\mu A^a_\nu- \partial_\mu A^a_\nu+g_sf^{abc} A^a_\mu A^a_\nu$
is the gluon-field strength tensor   and $g_s$ is the quark-gluon
coupling, so that $\alpha_s=g_s^2/(4\pi)$, with summation over the colour indices $i=1,2,3$ and  $a=1,...8$.
From Eq.~(\ref{eq:Lqcd}) one derives the basic elements of the QCD Feynman graphs:
quark and gluon propagators and quark-gluon, 3-gluon and 4-gluon 
vertices. In QCD,  a crucial role is played by quark-gluon loop diagrams 
generating  the effective scale-dependent 
coupling $\alpha_s(\mu)$.  As we know, it 
logarithmically decreases at large scales, $\mu\to \infty$, ({\it asymptotic freedom})
as illustrated in Fig.~\ref{fig:alphas}.
The perturbation theory in terms of Feynman diagrams of quark-gluon interactions is well defined only at large  energy/momentum transfers. 
Inversely, 
at small momenta (long distances), as shown in the same Fig.~\ref{fig:alphas},
the coupling grows. 
At momentum transfers smaller than a few hundred MeV 
the perturbation theory for quarks and gluons in QCD is senseless. 
An intrinsic scale $\Lambda_{QCD}\sim 200-300$ MeV emerges, the quarks, antiquarks
and gluons interact strongly. Moreover, they are only observable in a form of 
coulourless bound states - the hadrons, one of them is the 
$B$ meson.

Another important feature of QCD 
concerns the vacuum state which is not an ``empty space'' in this theory.
It contains fluctuating quark-antiquark and
gluon  fields with characteristic wave lengths of $O(1/\Lambda_{QCD})$. 
Averaged densities of these fields known 
as vacuum condensate densities play an important role in our
story. In fact, the most important role will be played by 
the quark condensate with a density parametrized 
as the vacuum average of the Lorentz- and colour-invariant 
local operator  $\langle 0 | \overline{q}^iq^i|0\rangle\equiv \langle \bar{q}q\rangle \neq 0 $, ($q=u,d,s$)
with dimension $d=3$. Let me remind you that
$\langle \bar{q}q\rangle\neq 0 $  reflects the spontaneous breaking of chiral symmetry
in QCD. 
One acquires a set of vacuum condensate densities with dimensions $d=3,4,5,..$ 
formed by all possible colourless Lorentz-invariant operators 
built from quark and gluon fields.
E.g., the $d=4$ operator formed from two gluon-field strengths yields the gluon condensate density $\langle 0 |(\alpha_s/\pi)G^a_{\mu\nu}G^{a\,\mu\nu} |0\rangle\equiv \langle GG\rangle \neq 0$. Importantly,
there is no $d=2$ condensate in QCD. A review on vacuum condensates can be found in \cite{Ioffe}.

Returning to the process of  $B$-meson annihilation, from the point of view of QCD 
it is important that the energy scale of  quark-gluon interactions
binding  $b$  and $\bar{u}$ inside $B$ is characterized by  
the mass difference between the meson 
($m_B\simeq 5.3$ GeV) and heavy $b$-quark: 
\begin{equation}
\bar{\Lambda}\sim m_B-m_b\sim 500-700~ \mbox{MeV}.
\label{eq:barL}
\end{equation}
To quantify the above estimate we literally take   
$m_b=4.6-4.8$ GeV, the so called ``pole'' quark mass. 
Important is that quarks and gluons inside the $B$ meson 
have energies $\leq \bar{\Lambda}$ and hence interact strongly.
At such scales no perturbative expansion in $\alpha_s(\bar{\Lambda}) $ 
is possible and QCD Feynman graphs cannot be used.
Moreover, in addition to "valence" quarks,  the partonic components with 
soft gluons and $\bar{q} q$ -pairs: 
\begin{equation}
|B^-\rangle =| b\bar{u}\rangle \oplus  | b\bar{u} G\rangle \oplus 
| b\bar{u}\bar{q}q\rangle \oplus \dots \,,
\end{equation}
are important in forming the complete ``wave function'' of the
hadronic state $|B \rangle$. We also have to keep in mind 
that the  QCD vacuum state $\langle 0|$ is   
populated by nonperturbative fluctuating quark-antiquark and gluon  fields.      
We conclude that for the  
hadronic matrix element $\langle 0 |\bar{u}\gamma^\mu \gamma_5 b |B\rangle \sim f_B$
there is no  solution in QCD within perturbation theory.

One possibility to calculate this matrix element is to use
a numerical simulation of QCD on the lattice. An impressive 
progress in this direction has been achieved in recent years.  
We will stay within continuum QCD and follow the 
method of QCD sum rules.

\subsection{Correlation function of heavy-light quark currents}

According to the original idea \cite{SVZ}, (see also one of the
first papers on this subject \cite{AE})  
we start from defining  a suitable correlation function: 
an object calculable  in QCD 
and  simultaneously related to the hadronic parameter $f_B$:
\begin{equation}
\Pi_{\mu\nu}(q)=\int d^4x~ e^{iqx} \langle 0 |T\{\bar{u}(x)\gamma_\mu\gamma_5 b(x)\,
\bar{b}(0)\gamma_\mu\gamma_5 u(0) \}|0\rangle\,.
\label{eq:corr}
\end{equation}
This is an amplitude of an 
emission and absorbtion of  the $b\overline{u} $ quark pair in the vacuum 
by the external current  $\bar{u}\gamma_\mu\gamma_5 b$  and its conjugate 
$\bar{b}\gamma_\mu\gamma_5 u$ with a 4-momentum $q$.
The $b\to u$ current is the same as in the hadronic matrix element
(\ref{eq:fBdef}) of the leptonic decay. 
To simplify the further derivation, it is convenient to 
deal with a Lorentz-invariant amplitude,
multiplying the above correlation function by the 4-momenta:
$q^\mu q^\nu\Pi_{\mu\nu}(q)\equiv \Pi_5(q^2)$. 
This is equivalent to taking divergences of 
the axial current operators under the $x$ integral:
$\partial^\mu(\bar{u}\gamma_\mu \gamma_5b)=(m_b+m_u)\bar{u}i\gamma_5b \equiv  j_5$
and replacing the axial currents by the 
pseudoscalar ones.
Hence, we may redefine the correlation function to a slightly different 
form 
\begin{equation}
\Pi_5(q^2)=\int d^4x~ e^{iqx} \langle 0 |T\{j_5(x) 
j^\dagger_5(0)\}|0\rangle\,,
\label{eq:corr5}
\end{equation}
so that $\Pi_5(q^2)$ depends only on the invariant  4-momentum square.
We accordingly modify the definition of the decay constant
\begin{equation}
p_{B}^\mu \langle 0|\bar{u}\gamma_\mu\gamma_5 b| B (p_B)\rangle=
\langle 0|j_5| B (p_B)\rangle = m_B^2 f_B \,.
\label{eq:fBdef2}
\end{equation}
Let us consider the correlation function (\ref{eq:corr5}) in the region $q^2\ll m_b^2$.
In the rest frame, $\vec{q}=0$, $q^2=q_0^2$ and the energy deficit to produce a real $B$ meson state 
from the current is $\Delta q_0=m_B-q_0\sim m_b$, up to small corrections. Thus, the propagation 
of the $b\bar{u}$ pair emitted by the current   $j_5(x)$ and absorbed by the current $j^+_5(0)$ 
lasts a time interval $\Delta x_0\sim 1/\Delta q_0\sim 1/m_b$, much shorter than a 
time/distance interval $\Delta x_0\sim \Delta x_i\sim 1/\Lambda_{QCD}$ typical for
the nonperturbative, strong interaction regime of QCD.
Hence the quark-antiquark pair propagation described by the correlation function
at $q^2\ll m_b^2$ remains highly virtual and therefore calculable in perturbative QCD.

\begin{figure}[h]
\centerline{
\includegraphics[width=0.49\textwidth]{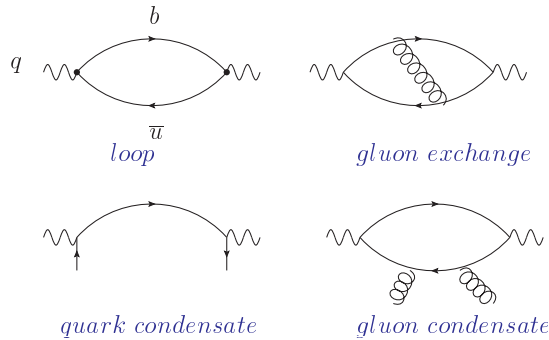}\hspace{1cm}}
\caption{\small Diagrams corresponding to the correlation function (\ref{eq:corr5}):
simple quark-antiquark loop diagram (upper left), one of the perturbative 
gluon-exchanges (upper right), quark condensate (lower left) and 
one of the gluon condensate diagrams (lower right).}
\label{fig:corrdiags}
\end{figure}

In the leading order of perturbation theory,  the function  $\Pi_{5}(q^2) $ is determined 
by a simple quark-loop diagram
 shown in Fig.~\ref{fig:corrdiags} (upper left).  Gluon radiative corrections 
to this diagram, one of them shown in Fig.~\ref{fig:corrdiags} (upper right)  
are suppressed  by small coupling $\alpha_s(\mu\sim m_b)$.
The simple loop diagram and radiative gluon corrections 
expressed via two- and three-loop diagrams (the latter were calculated in \cite{ChetS}) form 
the perturbative part of the correlation  function $\Pi_5(q^2)$. 

Additional diagrams shown in Fig~\ref{fig:corrdiags} 
take into account  the interactions with 
QCD vacuum fields. A detailed calculation of the quark condensate
diagram shown in Fig.~\ref{fig:corrdiags} (lower left) can be found e.g., in the review \cite{KR98}. 
The gluon condensate diagrams (one of them in Fig~\ref{fig:corrdiags}  (lower right)) 
are more complicated because they represent a combination of the loop and vacuum insertions.
Useful methods to calculate these diagrams are introduced in the review \cite{technrev}.
Technically, one uses Feynman rules of QCD and considers the vacuum
quark-antiquark pairs and gluons as external static fields.
There are also contributions combining the quark-antiquark and gluon vacuum lines.
All condensate diagrams forming the nonperturbative part of $\Pi_5(q^2)$ 
and calculated at $q^2\ll m_b^2$ contain a short-distance
part, formed by the propagating  quarks and antiquarks, and a long-distance
part approximated by averaged condensate densities. This is how  a short-distance 
quark-antiquark fluctuation ``feels'' the QCD vacuum, ``taking the snapshots'' \cite{Shifman98}
of it.
   
The result for the correlation $\Pi_5(q^2)$ 
is  an analytical expression in terms of  the quark masses $m_b$, $m_u$, quark-gluon coupling $\alpha_s$  and universal QCD condensate densities.
Interpreting the calculational procedure as a systematic  OPE
is another important theoretical aspect. An introduction
to the OPE adapted for the correlation functions in the presence of 
vacuum condensates can be found e.g., in \cite{Shifman98,CK}. Formally, one expands
the product of two current operators in a series of local operators with growing dimensions,
built from quark, antiquark fields and gluon field strength:
\begin{equation}
T\{j_5(x) j_5^\dagger(0)\}= \!\!\!\sum\limits_{d=0,3,4,..} C_d(x^2,m_b,m_u,\alpha_s) O_d(0)\,.
\label{eq:OPE}
\end{equation}

Taking vacuum average of the above formula and integrating 
it over $x$ we recover  the correlation function:  
\begin{equation}
\Pi_5(q^2)=\int d^4x\, e^{iqx}
\langle 0 |T\{j_5(x) j_5^\dagger(0)\}|0\rangle
=\sum\limits_{d=0,3,4,..} \overline{C}_d(q^2,m_b,m_u,\alpha_s)
\langle 0 |O_d|0\rangle\,,
\label{eq:ope2}
\end{equation}
where $\overline{C}_d(q^2,...)=\int d^4x e^{iqx}C_d(x^2,...)$. 
Evidently, only the operators
with vacuum quantum numbers 
(Lorentz-scalar, $C$-, $P$-, $T$-invariant, colourless) contribute to the r.h.s : 
\be
O_0=1, ~~O_3=\bar{q}q, ~~O_4=G^{a}_{\mu\nu}G^{a\mu\nu} , ~~
O_5=\bar{q}\sigma_{\mu\nu}\frac{\lambda^a}2 G^{a}_{\mu\nu}q,~~
O_6=(\bar{q}\Gamma_r q)((\bar{q}\Gamma_r q),...~~,
\label{eq:oper}
\ee
where $q=u,d,s$,  and $\Gamma_r$ are certain combinations of  Dirac matrices. 
The unit operator
with $\langle 0 |O_0|0\rangle=1$ and no fields is added for the sake of uniformity.
Its coefficient represents the perturbative part of the correlation
function, $\Pi_5^{(pert)}(q^2)=\overline{C}_0(q^2)$.
This part is obtained from the loop diagram and gluon radiative corrections and is
conveniently  represented in the form of a dispersion integral:  
\be
\Pi_5^{(pert)}(q^2)-\Pi_5^{(pert)}(0)-q^2 \frac{d}{dq^2}\Pi_5^{(pert)}(0)=(q^2)^2\int \limits_{m_b^2}^\infty\! ds \,
\frac{\rho_{5}^{(pert)}(s)}{s^2(s-q^2)}
\label{eq:Pi5pert}
\ee
with the spectral density 
\be
\rho_{5}^{(pert)}(s)=\frac1{\pi}\mbox{Im}\Pi^{(pert)}_5(s)=\frac{3 m_b^2}{8\pi^2}s\left(1-\frac{m_b^2}{s}\right)^2
+ O(\alpha_s)  + O(\alpha^2_s)\,.
\label{eq:pertLO5}
\ee
The two subtractions are needed for  the convergence of the integral. 
Note that for simplicity we neglected the light-quark mass in Eq.~(\ref{eq:pertLO5}).  
The $O(\alpha_s)$ and $O(\alpha_s)^2$ corrections in this equation  are 
considerably more complicated and can be found in \cite{ChetS,JL} (see also e.g., \cite{GKPR}).

The dominant nonperturbative contribution to the OPE (\ref{eq:ope2}) 
stems from the quark condensate:
 \ba
\label{PSqq}
 \Pi^{\langle \bar q q\rangle}_5 (q^2) =  \overline{C}_3(q^2)\langle \bar q q\rangle,
~~~~~\mbox{where}~~
\overline{C}_3(q^2)=\frac{-m_b^3}{m_b^2-q^2}+O(\alpha_s)\,.
\ea
The leading order result for the Wilson coefficient $\overline{C}_3(q^2)$
is obtained from the diagram shown in Fig.~\ref{fig:corrdiags} (lower left) 
and a more complicated
expression for the $O(\alpha_s)$  gluon radiative correction can be found in \cite{GKPR}. 
In the above expression, the separation of  short and long distances
is visible:  the short-distance part is given by a simple $b$-quark propagator 
with 4-momentum $q$  whereas the quark condensate density 
represents the long-distance effect.
The complete expression for the correlation function 
in a compact form is:
\be
\Pi_5^{(OPE)}(q^2)=\Pi_5^{(pert)}(q^2)+\Pi_5^{\langle \bar q q\rangle}(q^2)+
\Pi_5^{\langle d456\rangle}(q^2)\,,
\label{eq:opeexp}
\ee
where all $d=4,5,6$ effects are collected in one term for brevity.
The terms with $d>6 $  are usually neglected, provided one 
keeps the $d=4,5,6$ contribution sufficiently small, due to a proper choice of 
the variable $q^2$.

\subsection{Correlation function in terms of hadrons}

\begin{figure}[h]
\begin{center}
\includegraphics[width=0.4\textwidth]{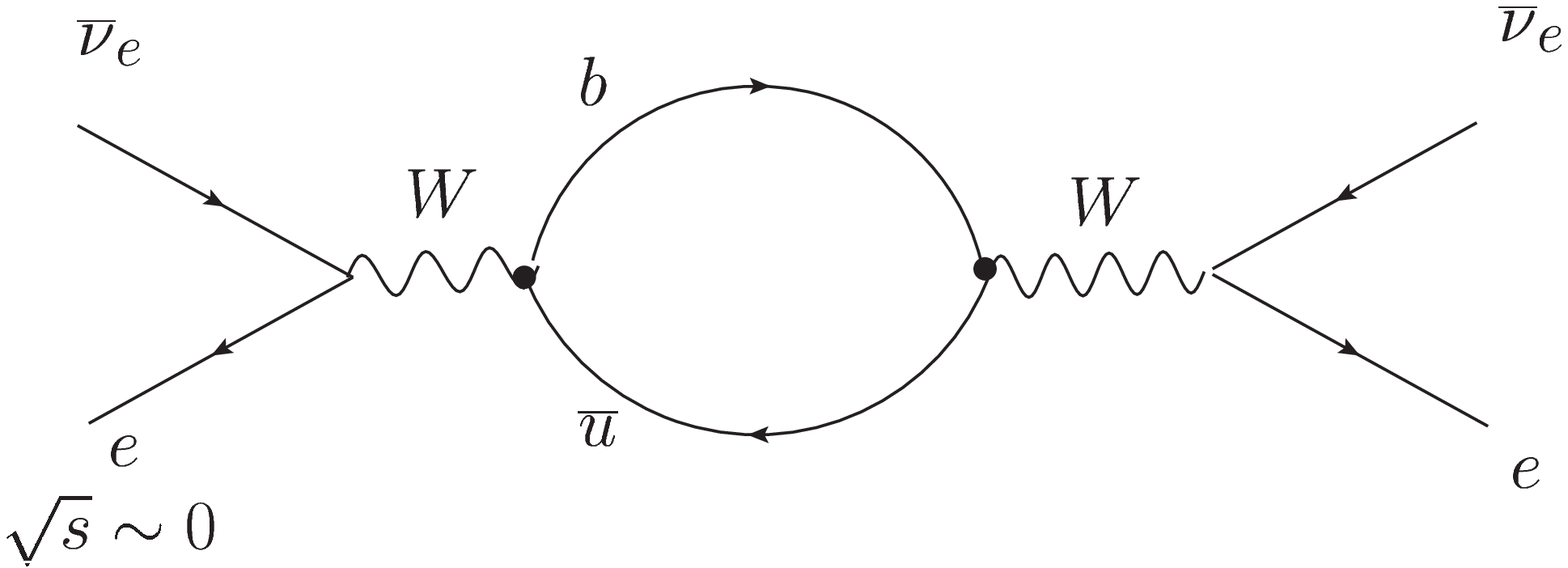}~~~
\includegraphics[width=0.4\textwidth]{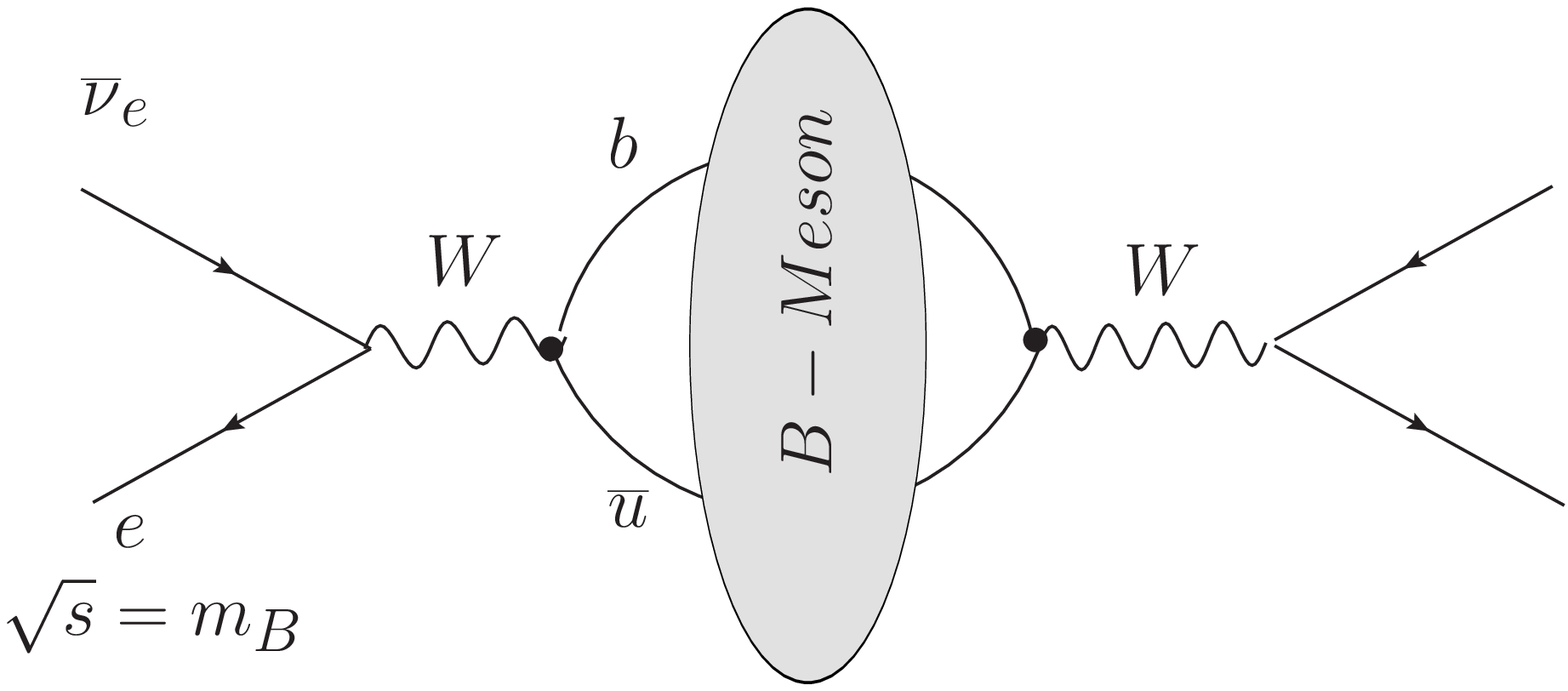} \\
{\small (a) \hspace{5cm} (b)}
\\[5mm]
\end{center}
\begin{center}
\includegraphics[width=0.4\textwidth]{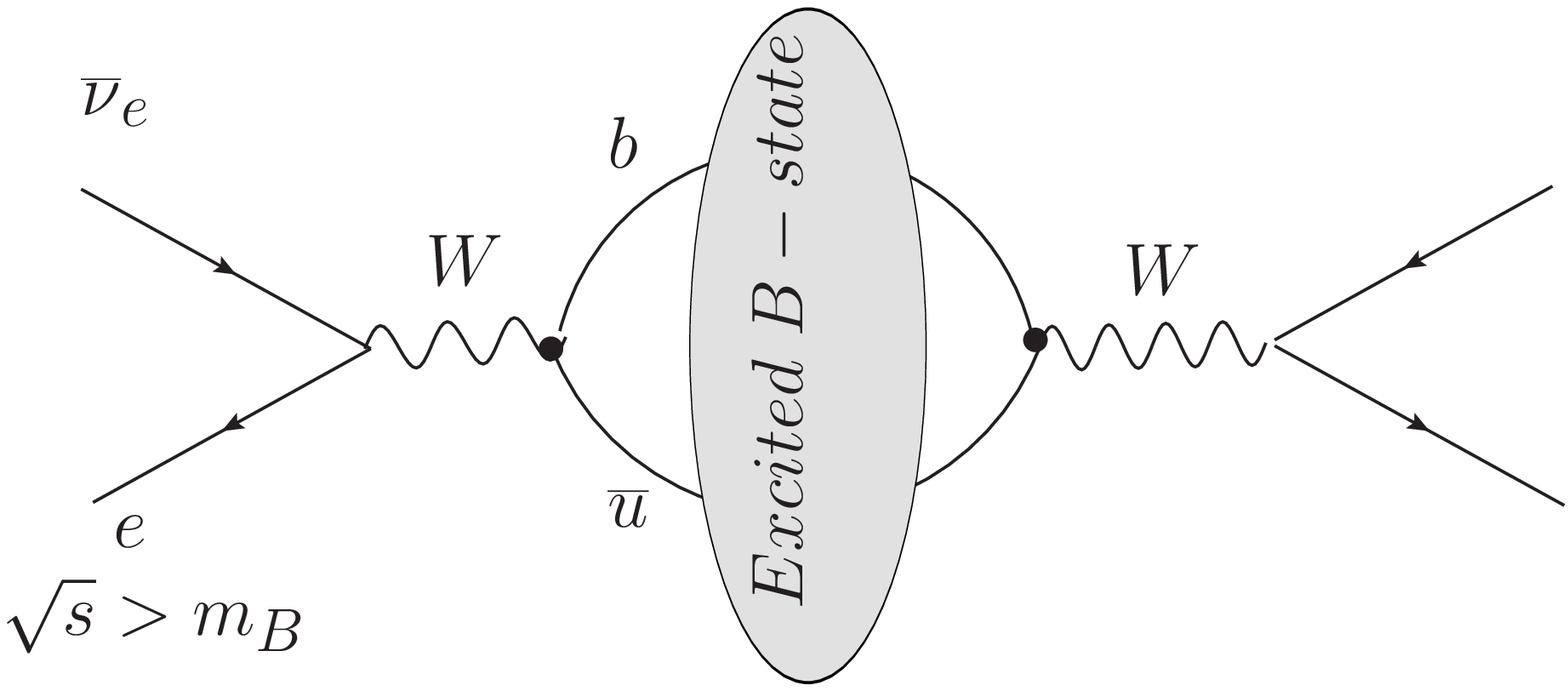}~~~
\includegraphics[width=0.4\textwidth]{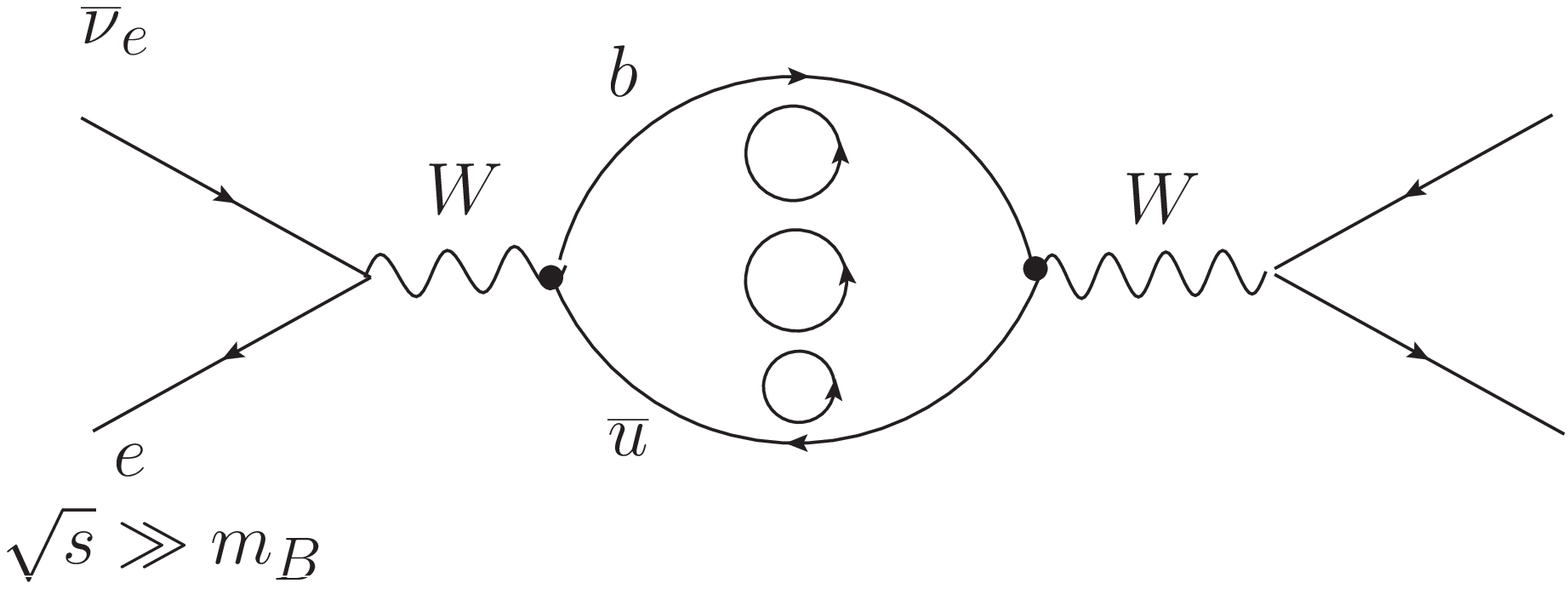}\\
{\small (c) \hspace{5cm} (d)}
\end{center}
\caption{ \small Correlation function as a part of the 
$\bar{\nu}_e e$ scattering amplitude at  different energies $\sqrt{s}=\sqrt{q^2}$.}
\label{fig:disp}
\end{figure}

Having at hand the expression (\ref{eq:opeexp}) for the correlation function
$\Pi_5(q^2)$ valid at  $q^2\ll m_b^2$, let us now 
investigate its relation to hadrons.  
To visualize the discussion, I consider a hypothetical neutrino-electron 
elastic scattering via a virtual  
$W$ boson. One of the possible intermediate states  
in this process is the $b\bar{u}$  pair emitted from and annihilated 
into $W$  (in the longitudinal state, to have $J^P=0^-$)
as depicted in Fig.~\ref{fig:disp}. The $b\bar{u}$ fluctuation 
coincides with the correlation function we are considering.
The c.m. energy of this process is equal to the momentum transfer 
in the correlation function: $\sqrt{s}=\sqrt{q^2}$.
In the region $q^2\ll m_b^2$ the intermediate $b\bar{u}$ 
state (Fig.\ref{fig:disp}(a)) represents a 
highly virtual heavy-light quark-antiquark pair. We are able to calculate 
this fluctuation in terms of OPE as already explained.
On the other hand, increasing the energy one reaches the domain where 
real on-shell hadronic states 
propagate  in the intermediate state. At $\sqrt{s}=m_B$, 
the $B$-meson (Fig.\ref{fig:disp}(b)) contributes. 
This is the lowest possible intermediate hadronic state  in this channel, 
it will show up  as a sharp resonance in  our hypothetical 
scattering process. Increasing the energy, one encounters 
heavier resonances, the radially excited  $B$ mesons,  with growing total width 
(Fig.\ref{fig:disp}(c)). These resonances are overlapped
with multihadron states  with a net $B$ flavor (Fig.\ref{fig:disp}(d)), starting  
with the two-particle hadronic state $B^*\pi$ with the lowest
threshold $s=q^2=(m_{B^*}+m_\pi)^2$. Note that a $B\pi$ state is not allowed 
by spin-parity conservation. The multihadron  state contributions build up
the hadronic continuum mixed with excited states.
At very large energies, resonances are smeared  and  multihadron states
dominate.
We come to conclusion that the correlation function $\Pi_5(q^2)$  
in the region $q^2> m_B^2$ describes 
a complicated overlap of interfering resonant and continuum 
hadronic states with $B$ meson quantum numbers.

This qualitative picture of emerging  
intermediate hadronic states reflects 
the formal spectral representation of $\Pi_5(q^2)$  
following from  the basic unitarity relation.
The imaginary part
of the correlation function
is equal to the sum of contributions of all
possible hadronic states allowed by quantum numbers: 
\begin{equation}
\frac{1}{\pi}\mbox{Im}\Pi_{5}(q^2)=\langle 0 |j_5 |B\rangle
\langle B | j_5^{\dagger}|0\rangle \delta(m_B^2-q^2)+
\rho^h_5(s)\theta(s-(m_{B^*}+m_\pi)^2)\,,
\label{eq:spectr}
\end{equation}
where we isolated the ground-state $B$ meson contribution
and introduce a shorthand notation for the spectral density 
of excited (resonance and  multiparticle) states, schematically:
\be
\rho^h_5(s)=\sum_{B_{exc}}
\langle 0 |j _5|B_{exc}\rangle\langle B_{exc} |j_5^{\dagger}|0\rangle
\label{eq:rhoh}
\ee
where the sum includes the integration over phase space and sum over polarizations.

The next important step is to employ the analyticity 
of the function $\Pi(q^2)$ which, according to the unitarity relation 
(\ref{eq:spectr}) has singularities -- poles (cuts) related to resonances 
(multiparticle thresholds) -- on the real positive axis of the complex $q^2$
plane. The Cauchy theorem leads  to the dispersion relation between 
$\Pi_5(q^2)$ and its imaginary part integrated over positive $s\geq m_B^2$:
\be
\Pi_5(q^2)=\frac{1}{\pi}\int\limits_{m_B^2}^{\infty}
ds \frac{\mbox{Im} \Pi_5(s)}{s-q^2-i\epsilon }\,,
\label{eq:disprel1}
\ee
where the subtraction terms are hereafter neglected for simplicity. 
Importantly, this relation is valid at any $q^2$.
We will apply it at $q^2\ll m_b^2$ where the correlation function represents 
a short-lived $b \bar{u}$ -fluctuation
calculable in terms of OPE, so that l.h.s. in the above dispersion relation 
can be approximated by $\Pi_5^{(OPE)}(q^2)$ given by Eq.~(\ref{eq:opeexp}).
Hence, we obtain a remarkable opportunity to relate the  
correlation function calculated in QCD to a sum/integral containing 
hadronic parameters, including the $B$-meson mass and decay constant.

\subsection{Deriving the sum rule for $f_B^2$}

Substituting Eq.~(\ref{eq:spectr}) in the dispersion relation (\ref{eq:disprel1}) 
and expressing  the hadronic matrix element via $f_B$,
we obtain at $q^2\ll m_b^2$:
\begin{equation}
\Pi_5(q^2)=\frac{f_B^2 m_B^4}{m_B^2-q^2}+\int\limits_{s_h}^{\infty}ds 
\frac{\rho^h(s)}{s-q^2} \simeq \Pi^{(OPE)}_5(q^2)\,.
\label{eq:disphadr}
\end{equation}
where $s_h=(m_{B^*}+m_\pi)^2$ is the lowest threshold of the excited $B$ states.

Let us now employ another important feature
of the correlation function. In the deep spacelike region $q^2\to -\infty$ 
the power suppressed condensate terms   
in Eq.~(\ref{eq:opeexp}) vanish and the 
correlation function coincides with the perturbative part of OPE: 
\be
\Pi_5(q^2\to -\infty) =\Pi^{(OPE)}_5(q^2\to -\infty)=\Pi^{(pert)}_5(q^2\to -\infty)\,,
\label{eq:ope4}
\ee 
dominated by the simple loop diagram.

It is convenient to express the perturbative part of the 
OPE in a form of dispersion relation.
In this case the imaginary part starts at the $b\bar{u}$-quark pair threshold 
and is equal to the spectral density of the loop diagrams
presented in Eq.~(\ref{eq:pertLO5}): 
\begin{equation}
\Pi^{(pert)}(q^2)=\frac{1}{\pi}\int\limits_{m_{b}^2}^{\infty}
ds \frac{\mbox{Im} \Pi^{(pert)}_5(s)}{s-q^2}\,,
\label{eq:dispOPE}
\end{equation}
where we again neglect the subtractions and put $m_u\to 0$.

To fulfill the asymptotic condition (\ref{eq:ope4}), the spectral functions
entering the  hadronic and OPE (perturbative) dispersion relations should 
be equal at sufficiently large $s$: 
\begin{equation}
\rho^{h}(s)\simeq\frac{1}{\pi}\mbox{Im} \Pi_5^{(pert)}(s),  
\label{eq:dual1}
\end {equation}
This approximation is called  local quark-hadron duality. 
It suffices to use a weaker condition,
approximately equating the integrals of   
the hadronic and perturbative spectral densities 
over the large $s$ region: 
\begin{equation}
\int\limits_{s_h}^{\infty}ds 
\frac{\rho^h(s)}{s-q^2}\simeq
\frac{1}{\pi}\int\limits_{s_0}^{\infty} 
ds \frac{\mbox{Im} \Pi^{(pert)}_5(s)}{s-q^2}\,,
\label{eq:dual2}
\end {equation}
where an  effective threshold $s_0$ is introduced.
Returning to the hadronic dispersion relation (\ref{eq:disphadr}),  we use Eq.~(\ref{eq:dual2})
to replace the integral over excited $B$ states  in l.h.s., and  use the OPE (\ref{eq:opeexp}) in r.h.s.,  with the perturbative part replaced by its dispersion representation. 
The resulting relation: 
\begin{equation}
\frac{f_B^2m_B^4}{m_B^2-q^2}+\frac{1}{\pi}\int\limits^{\infty}_{s_0}
ds \frac{\mbox{Im} \Pi^{(pert)}_5(s)}{s-q^2}=
\frac{1}{\pi}\int\limits^{\infty}_{m_b^2}
ds \frac{\mbox{Im} \Pi^{(pert)}_5(s)}{s-q^2}
+\Pi_5^{\langle \bar q q\rangle}(q^2)+
\Pi_5^{\langle d456\rangle}(q^2)\,,
\label{eq:OPE3}
\end{equation}
allows one to subtract the approximately equal integrals from both sides
yielding an analytical relation for the decay constant:
\begin{equation}
\frac{f_B^2m_B^4}{m_B^2-q^2}
=\frac{1}{\pi}\int\limits^{s_0}_{m_b^2}
ds \frac{\mbox{Im} \Pi^{(pert)}_5(s)}{s-q^2}
+\Pi_5^{\langle \bar q q\rangle}(q^2)+
\Pi_5^{\langle d456\rangle}(q^2)\,.
\label{eq:OPE4}
\end{equation}
 A substantial improvement of this relation is further achieved 
with the help of the Borel transformation defined as: 
\begin{equation}
\Pi_5(M^2) \equiv 
{\cal B}_{M^2}\Pi_5(q^2)=\lim_{\stackrel{-q^2,n \to \infty}{-q^2/n=M^2}}
\frac{(-q^2)^{(n+1)}}{n!}\left( \frac{d}{dq^2}\right)^n \Pi_5(q^2)~,
\label{eq:B}
\end{equation}
so that $ {\cal B}_{M^2}(\frac{1}{m^2-q^2}) = \exp(-m^2/M^2)$. 

The resulting QCD sum rule  for $f_B^2$ obtained from Eq.~(\ref{eq:OPE4}) 
after this transformation  reads:
\begin{equation}
f_B^2m_B^4e^{-m_B^2/M^2}=\int\limits_{m_b^2}^{s_0} ds e^{-s/M^2}
\mbox{Im}\Pi_5^{(pert)}(s,m_b,m_u,\alpha_s) 
+\Pi_5^{\langle \bar q q\rangle}(M^2)+
\Pi_5^{\langle d456\rangle}(M^2)\,.
\label{eq:fBSR}
\end{equation}
Note that the Borel transformation suppresses the higher-state contributions 
to the hadronic sum above $s_0$ so that the above sum rule 
is less sensitive to the accuracy of the quark-hadron duality approximation
(\ref{eq:dual2}).
Everything is ready to calculate the decay constant of $B$ meson numerically.

\subsection{Input parameters and results}
In the sum rule  (\ref{eq:fBSR}) one has to choose an optimal interval 
of the Borel  parameter. The lower boundary for $M^2$ is controlled by the OPE
convergence, e.g., we  demand that the $d=4,5,6$ terms are sufficiently 
small with respect to the quark condensate  term.  The upper boundary 
for $M^2$ is adopted from the condition that the contribution of excited states
subtracted from the sum rules remains subdominant.
Furthermore, a  standard way to fix the effective parameter $s_0$ is 
to fit the sum rule to the measured mass of $B$-meson 
by differentiating both parts of Eq.~(\ref{eq:fBSR}) in $-(1/M^2)$
and dividing the result by the initial sum rule, so that $f_B^2$ cancels,
and one  obtains  a relation for $m_B^2$.

One of the advantages of the sum rule method is its flexibility:  replacing
quark flavours in the correlation function, e.g., $b \to c$ or $\bar{u}\to \bar{s}$ 
provides an access to the decay constants of $D$ or $B_s$ mesons.
\begin{table}[h]
\centerline{\begin{tabular}{|c|c|c|}
\hline
 Decay constant & Lattice QCD [ref.] & QCD sum rules \cite{GKPR}\\
\hline
& 196.9 $\pm$ 9.1 \cite{Fermilab}& \\[-1mm]
$f_{B}$[MeV] & & $207^{+17}_{-9}$\\[-1mm]
& 186 $\pm$ 4 \cite{HPQCDb} & \\
\hline
& 242.0 $\pm$ 10.0 \cite{Fermilab}  &\\[-1mm]
$f_{B_s}$[MeV] & &$242^{+17}_{-12}$\\[-1mm]
& 224 $\pm $ 5 \cite{HPQCDb} &\\
\hline
&1.229$\pm$ 0.026  \cite{Fermilab}   &\\[-1mm]
$f_{B_s}/f_B$ & & $1.17^{+0.04}_{-0.03}$\\[-1mm]
&1.205$\pm$ 0.007 \cite{HPQCDb}  &\\
\hline\hline
& 218.9 $\pm$ 11.3  \cite{Fermilab} &\\[-1mm]
$f_D$[MeV] & &$201^{+12}_{-13}$\\[-1mm]
& 213 $\pm$ 4 \cite{HPQCDc}&\\
\hline
& 260.1 $\pm$ 10.8  \cite{Fermilab}&\\[-1mm]
$f_{D_s}$[MeV]& &$238^{+13}_{-23}$\\[-1mm]
 &248.0 $\pm$ 2.5 \cite{HPQCDc} &\\
\hline
&1.188$\pm$ 0.025   \cite{Fermilab}  &\\[-1mm]
$f_{D_s}/f_D$  &&$1.15^{+0.04}_{-0.05}$\\[-1mm]
&1.164$\pm$ 0.018 \cite{HPQCDc}&\\
\hline
\end{tabular}}
\caption{ \small Decay constants of heavy-light mesons calculated with different methods.}
\label{tab:fBfD}
\end{table}
Nonzero strange quark mass and a difference in  condensate densities,
$\langle \bar{s}s\rangle \neq \langle \bar{u}u \rangle $, 
generate the $SU(3)_{flavour}$ symmetry violation. 

The universal input parameters needed for the  numerical analysis of 
the sum rules include the quark masses, quark-gluon coupling and the 
vacuum condensate  densities. Since the calculation is done at short
distances, the natural choice for quark masses is the  
$\overline{MS}$ scheme. The sum rule  is quite sensitive to the $b$-quark mass,
hence to have a reliable  estimate of $f_B$ one needs   
an independent and accurate determination of $m_b$.
This task was fulfilled by considering  quarkonium sum rules, 
where the correlation function of two $\bar{Q}\gamma_\mu Q$ 
currents ($Q=b,c$) is calculated in QCD. The accuracy 
of this calculation \cite{mbmcChet_etal} has reached  $O(\alpha_s^3)$ 
in the perturbative part.  The hadronic representation of this correlation function 
is largely fixed from experiment \cite{PDG} and  
consists of  $J^{PC}=1^{--}$ heavy quarkonia  levels, their decay constants 
measured in $e^+e^-\to \Upsilon,\Upsilon(2S),....$  or $e^+e^-\to J/\psi,\psi(2S),....$.
Hence, the quarkonium sum rules can be used to extract the heavy quark masses.
The most recent results of these determinations \cite{mbmcChet_etal,Hoangmc}, expressed   
in $\overline{MS}$ scheme are very close to the PDG averages:
$\bar{m}_b(\bar{m}_b)= (4.18 \pm 0.03)\,\mbox{GeV}$, 
$\bar{m}_c(\bar{m}_c)= (1.275\pm 0.025)\,\mbox{GeV}$ \cite{PDG}.
In the same way,  employing QCD sum rules for strange meson pseudoscalar and scalar
channels \cite{ms} one determines $m_s$ consistent with
$m_s(\mu=2\,\mbox{GeV}) = (95 \pm 10)\,\mbox{MeV}$\cite{PDG}. Combining $m_s$ 
with ChPT  relations \cite{Leutwyler} one finds the   quark condensate density 
$\langle \bar{q}q\rangle(\mbox{2 GeV})= - (277^{+12}_{-10} ~\mbox{MeV})^3$.
Condensate densities with $d>3$ entering the subleading power corrections in OPE 
are mainly taken from the review \cite{Ioffe}.  The recent  determinations
of the $B$ and $D$ decay constants in Table~\ref{tab:fBfD} are taken from \cite{GKPR} 
where one can also find  a detailed discussion of numerical procedure and 
formulae for OPE, as well as references to other important papers on the subject
of this lecture.

\section{\!\!\!\!\!\!.~Lecture: 
$B\to \pi$  form factors 
and light-cone sum rules}

In this lecture more complicated hadronic matrix elements --
the form factors of heavy-to-light transitions are considered. The best studied 
among them are the $B\to \pi$ transition form factors 
relevant for   $B\to \pi \ell \nu_\ell$
semileptonic decay. I will explain how the QCD sum rule method was modified
to calculate these and other hadronic form factors.

\subsection{ $B\to \pi \ell \nu_\ell$ decay  and 
form factors }
\begin{figure}[hb]
\centerline{
\includegraphics[width=0.45\textwidth]{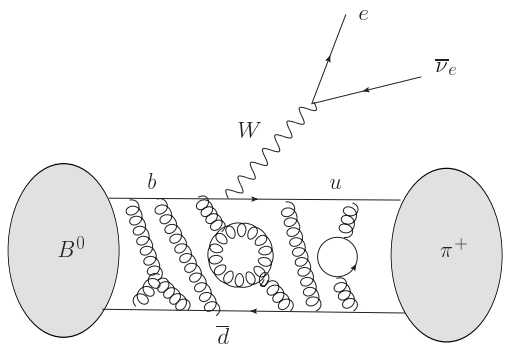}\hspace{1cm}
}
\caption{\small Schematic view of $B\to \pi \ell \nu_\ell$ decay}
\label{fig:Bpiff}
\end{figure}
The exclusive semileptonic decay $\bar{B}^0\to \pi^+\ell \bar{\nu}_\ell$ shown 
in Fig.~\ref{fig:Bpiff} proceeds via weak $b\to u$ transition with a 
squared momentum transfer $q^2$ to the leptonic pair varying within the interval
$0<q^2<(m_B-m_{\pi})^2\sim 26~ \mbox{GeV}^2$ (here we neglect the lepton mass).  

The form factors  $f^+_{B\pi}(q^2)$ and $f^0_{B\pi}(q^2)$ are invariant functions of $q^2$ parameterizing  the hadronic matrix element of this decay:
\begin{eqnarray} 
\langle \pi^+(p)|\bar{u} 
\gamma_\mu   b |\bar{B}^0(p+q)\rangle
=f^+_{B\pi}(q^2)\Big [2p_\mu +
\big (1-\frac{m_B^2-m_\pi^2}{q^2}\big) q_\mu\Big] 
+ f^0_{B\pi}(q^2)\frac{m_B^2-m_\pi^2}{q^2}q_\mu,
\label{eq:Bpihme}
\end{eqnarray} 
where $p+q$ and $p$ are the four-momenta of $B$ and $\pi$, respectively.
Similar to the $B$ decay constant, the $B\to \pi$ form factors
have to be calculated in QCD. This is a challenging 
problem because not only the initial $B$ meson but also the 
final pion is involved in the hadronic matrix element. In what follows, we 
consider the region of small $q^2$, in which case the pion has a large recoil in the 
$B$ meson rest system,  with the momentum $p_\pi\equiv |\vec{p}\,|\sim m_B/2$ 
at $q^2=0$. 

Analyzing the $B\to \pi$ form factors from the point of view of QCD,
one expects a certain perturbative contribution corresponding 
to an energetic virtual gluon exchange between the quarks 
participating in the weak transition and the spectator quark. This 
``hard scattering''  mechanism 
boosts the spectator quark in $B$ meson and provides a natural configuration
for the final pion with symmetric collinear quark and antiquark. On the other 
hand one has to take into account also the ``end-point'' mechanism 
where the pion is formed from an asymmetric quark-antiquark pair.
This part of the  form factor is dominated by soft nonperturbative gluons.
The proportion of the 
hard scattering and soft end-point contributions to the hadronic form factors   
is a long-standing problem.  It can only be addressed within a calculational method
that allows one to take into account  both contributions.

An accurate determination of the $B\to \pi$ form factors is important 
for quark flavour physics because the semileptonic decay $B\to \pi \ell \nu_\ell$ is 
an excellent source of the CKM parameter  $|V_{ub}|$. In fact, one  practically
needs only the vector form factor $f^+_{B\pi}$ for this purpose, because in the partial width 
the contribution of the form factor $f^0_{B\pi}$ is suppressed by the lepton mass:
\begin{eqnarray} 
\frac{1}{\tau_{B^0}}\frac{d BR(\bar{B}^0\to \pi^+l^-\nu)}{dq^2}=
\frac{G_F^2 |V_{ub}|^2}{24\pi^3}p_\pi^3
|f^+_{B\pi}(q^2)|^2 +O(m_l^2)\,. 
\label{eq:Bpilnu}
\end{eqnarray} 
Importantly, in lattice QCD the $B\to \pi$ form factors are 
currently accessible  at  comparatively large $q^2\geq 15$ GeV$^2$. 
In this region the phase space 
in the decay width (\ref{eq:Bpilnu}) is suppressed by small $p_\pi$.
The calculation of the form factors at small $q^2$ (large recoil of the pion) 
discussed below,  
complements the lattice QCD results in a kinematically dominant region.

\subsection{ Vacuum-to-pion correlation function}

\begin{figure}[b]
\centerline{
\includegraphics[width=0.75\textwidth]{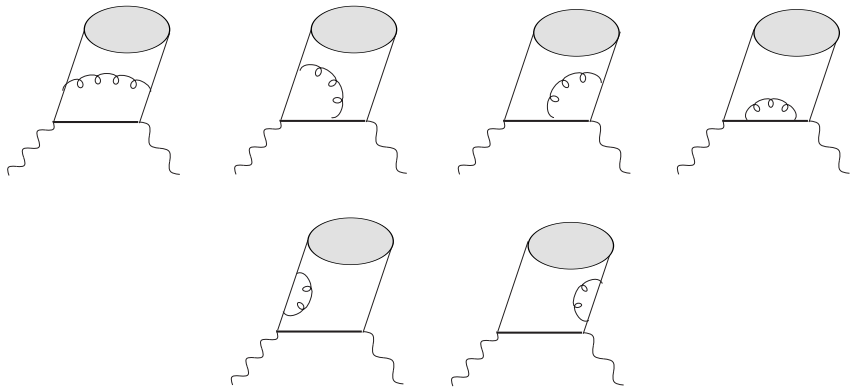}}
\caption{ \small Gluon radiative corrections to the correlation function (\ref{eq:lcsrcorr}). }
\label{fig:Bpicorrdiags2}
\end{figure}

The method of light-cone sum rules (LCSR) 
developed in \cite{lcsr1,Chernyak} is used to calculate the $B\to\pi$  form factors at large hadronic recoil. In this approach,  
the correlation function itself is an amplitude of the vacuum-to-hadron transition
\footnote{~
Vacuum-to-vacuum  correlation functions  
with the quark currents interpolating both $B$ meson and pion 
and with the OPE in terms of  condensates are 
not convenient for heavy-to-light form factors; see a detailed discussion 
in the review \cite{Braun_rev97}.}:
\begin{eqnarray}
F_\lambda (q,p)=
i \int d^4x~e^{iqx}\langle \pi(p)\mid T\{\bar{u}(x)\gamma_\lambda b(x), j^\dagger_5(0)\}\mid 0\rangle
\nonumber\\
= F(q^2,(p+q)^2)p_\mu+\tilde{F}(q^2,(p+q)^2)q_\mu\,,
\label{eq:lcsrcorr}
\end{eqnarray}
containing the product of the weak $b\to u$ and  $j_5=m_b\,\bar{b}i\gamma_5d$
currents. The latter
was also used in the two-point correlation function for $f_B$. 
In what follows, only the invariant amplitude $F$ is essential,
depending on the two independent kinematical variables: 
$ q^2$ , the squared momentum transfer in the weak $b\to u$ transition, 
and $ (p + q)^2$, the square of the 4-momentum 
flowing into  the current $j_5$. The correlation function (\ref{eq:lcsrcorr}) allows
for a systematic QCD calculation in the specific region:
$ q^2, (p + q)^2 \ll m_b^2 $ where the 
$b$ quark is a highly-virtual object. In this region of external momenta the 
$x$ integral in the correlation function is dominated
by small $x^2\sim 1/m_b^2$, near the light-cone $x^2\sim 0$.
The leading order diagram for the correlation function is shown in 
Fig.~\ref{fig:Bpicorrdiags1}(a). It consists of the 
free $b$-quark propagator convoluted with the 
matrix element of  light quark and antiquark operators
sandwiched between the vacuum and on-shell pion state.
The perturbative gluon corrections 
to the leading order diagram are shown in Fig.~\ref{fig:Bpicorrdiags2}.
The diagram in Fig.~\ref{fig:Bpicorrdiags1}(b)  takes into account
the emission of  a soft (low-virtuality) gluon emitted
from the $b$ quark. The corresponding  vacuum-pion matrix element
involves  light quark-antiquark and gluon fields.

A schematic expression for the correlation function
(\ref{eq:lcsrcorr}) decomposed near the light-cone can be written as:
\begin{figure}[t]
\begin{center}
\includegraphics[width=0.25\textwidth]{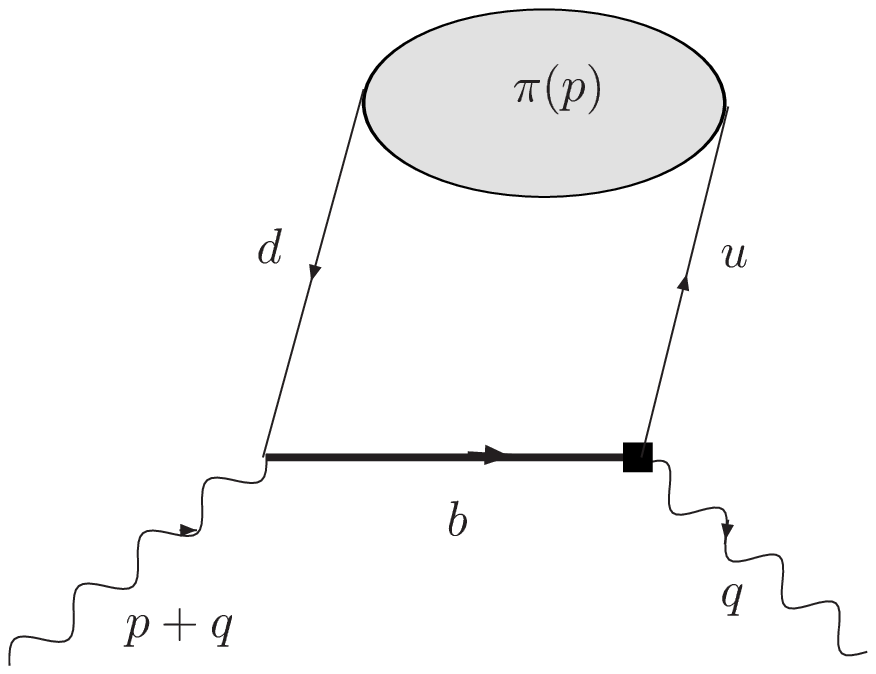}\hspace{1cm}~~~
\includegraphics[width=0.25\textwidth]{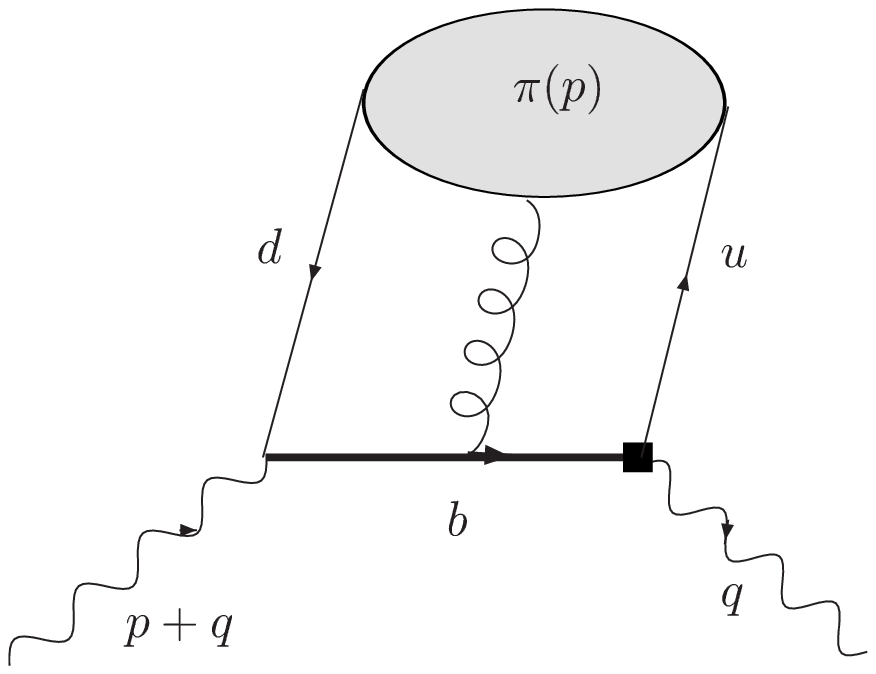}\hspace{1cm}\\
{\small (a) \hspace{4.5cm} (b)}
\end{center}
\caption{ \small Diagrams corresponding to the correlation function
(\ref{eq:lcsrcorr}):  leading order (a) and soft gluon emission forming the 
3-particle $B$ meson DA (b).}
\label{fig:Bpicorrdiags1}
\end{figure}
\begin{eqnarray}
\!\!\!\!\!F(q,p)=i \int d^4x\,e^{iqx}\Bigg\{ 
\left[ S^{0}(x^2,m_b^2)+\alpha_s S^{1}(x^2,m_b^2)\right]
\langle \pi(p)\mid \bar{u}(x)
\Gamma d(0)\!\mid \!0 \rangle
\nonumber  \\
+
\int_0^1 dv~\tilde{S}(x^2,m_b^2,v)
\langle\pi(p)\mid \bar{u}(x)G(vx)
\tilde{\Gamma}d(0)\}\mid 0\rangle
\Bigg\}
+...
\label{eq:lcexp}
\end{eqnarray}
where $S_{0}$ , $S_1$  and $\tilde{S}$  are the perturbative parts of the amplitudes,
involving $b$-quark propagators.  They are convoluted
 with the vacuum-pion matrix elements, taken  near $x^2=0$,
where $\Gamma,\tilde{\Gamma}$  are generic Dirac-matrix structures 
and the Lorentz-indices are omitted for simplicity.

The vacuum-pion matrix elements  in Eq.~(\ref{eq:lcexp}) 
are nonperturbative but universal objects.
They absorb all long-distance effects in the correlation function.
The expansion in Eq.(\ref{eq:lcexp}) goes over $\alpha_s$ and powers of $x^2$,
which in the momentum space translates into an expansion
in $\alpha_s(\mu)$ and the powers of $1/\mu$.
 Here   $\mu\sim \sqrt{\chi m_b}$ , with 
$\chi$  being  an intermediate scale, $\Lambda_{QCD} \ll \chi <m_b$.
In particular, in (\ref{eq:lcexp})
the quark-antiquark gluon part has a power suppression with respect to the leading 
order part.  Hence, the expansion (\ref{eq:lcexp}) can safely be truncated.  
I skip a more formal and systematic description of this expansion based on the twist  $t$ (dimension minus Lorentz-spin)
of the light  quark-antiquark operators entering the vacuum-pion matrix
elements (see, e.g., \cite{CK} for an introductory explanation).

The main nonperturbative object determining the leading-order answer 
for the light-cone expanded correlation function (\ref{eq:lcexp})  
is the vacuum-pion matrix element
\begin{equation}
\langle\pi(q)|\bar{u}(x)[x,0]\gamma_\mu\gamma_5 d(0)|0\rangle_{x^2=0}=
-iq_\mu f_\pi\int_0^1du\,e^{iuqx} \varphi_\pi (u) +O(x^2) ~,
\label{eq:pionDA}
\end{equation}
where the factor $[x,0]=exp[ig_s\int_0^1 dt x_\mu A^{a\mu}(tx)\lambda^a/2]$ is added to secure gauge invariance.
The above matrix element is normalized to the pion decay constant, which becomes evident 
if one puts $x\to 0$  and takes into account that the 
function $ \varphi_\pi (u) $ is normalized to unit. 
This and similar functions parameterizing 
vacuum-pion matrix elements play
the central role in the LCSR approach and replace 
the vacuum condensates. They are called light-cone distribution 
amplitudes (DA's) of the pion. Physically, DA's correspond
to various Fock components  of the pion and the variable $u$ 
in the two-particle DA (\ref{eq:pionDA}) denotes the share of the pion momentum carried 
by one of the constituents. 

Inserting in Eq.~(\ref{eq:lcexp}) the vacuum-pion matrix elements expressed 
in terms of DA's  and integrating over $x$,  one obtains 
the OPE result  for the invariant amplitude defined in (\ref{eq:lcsrcorr}) in the following generic form: 
\begin{equation}
F^{(OPE)}(q^2, (p\!+\!q)^2)
=\sum\limits_{t=2,3,4,..}\int du~
 T^{(t)}(q^2,(p\!+\!q)^2,m_b^2,\alpha_s,u,\mu)
\,\varphi^{(t)}_\pi(u,\mu)\,,
\label{eq:opelc}
\end{equation}
where the summation goes over the growing twist, and the twist-2 part 
contains the DA defined in Eq.~(\ref{eq:pionDA}).
The perturbative hard-scattering amplitudes $T^{(t)}$ stemming from the $b$-quark
propagators and perturbative loops are process-dependent 
whereas the pion DA's are universal.  One can analyse DA's
using the light-cone OPE  for other processes, not even involving 
heavy quarks, like  e.g., the pion electromagnetic form factor at spacelike 
momentum transfers or the photon-pion transition form factor (see e.g. \cite{CK}). 
Within the currently achieved accuracy, the light-cone OPE (\ref{eq:opelc}) 
includes  the twist 2,3,4 quark-antiquark and quark-antiquark-gluon DA's
\cite{BKR}, and the hard-scattering amplitudes for twist 2,3 parts are calculated up to NLO, in $O(\alpha_s)$ \cite{KRWY,Baganetal,BZ,DKMMO}. Recently, the 
$O(\alpha_s^2)$  correction to the twist-2 part was also  
calculated \cite{Bharucha}.

\subsection{What do we know about the light-cone DA's}

Before applying  them in LCSR's, the pion DA's were already introduced 
in the context of the hard-scattering mechanism for the pion
e.m. form factor at large momentum transfer \cite{ER,BL}. 
A convenient expansion in Gegenbauer polynomials  was defined
 \begin{equation}
\varphi_\pi(u,\mu)=6u(1-u)\left[1+\sum\limits_{n=2,4,..} a_{n}^\pi(\mu)
C_{n}^{3/2}(2u-1)\right],
\label{eq:Gegenb}
\end{equation}
with logarithmic evolution of its coefficients (Gegenbauer moments):
\begin{equation}
a_{2n}^\pi(\mu)\sim [\ln(\mu/\Lambda_{QCD})]^{-\gamma_{2n}}\,,
\label{eq:evol}
\end{equation}
vanishing at asymptotically large 
scale $\mu\to \infty$. The input values of Gegenbauer moments 
at low scale,   $a_{2,4,6,...}^\pi(\mu\sim\mbox{ 1 GeV})$ 
are determined from different sources: 
matching experimentally measured pion form factors to LCSR's , 
calculating  $a_{2}$ from two-point QCD sum rules  
and  in lattice QCD.
Recent determinations lie within  
the intervals:
$a_2^{\pi}=0.25\pm 0.15$,~ $a^{\pi}_2+a_4^{\pi}=0.1\pm 0.1$,
if one neglects  the higher coefficients.
The remaining parameters  of twist 3,4 DA's are mainly  determined 
from dedicated two-point sum rules \cite{BBL06}.

\subsection{LCSR for $B\to \pi$ form factors}
After obtaining the OPE expression for the amplitude $F((p+q)^2,q^2)$,
the derivation of LCSR follows the same strategy  as in the 
case of two-point sum rule. 
The hadronic dispersion 
relation  for $F((p+q)^2,q^2)$ in the variable $(p+q)^2$  and at fixed small $q^2$
is used.
The dispersion relation contains a pole term with intermediate $B$ meson
and a hadronic sum over excited and multihadron states with $B$ quantum numbers.
Matching the OPE with this dispersion relation, we obtain 
\begin{equation}
F^{(OPE)}((p+q)^2,q^2)= \frac{2m_B^2f_B f_{B\pi}^+(q^2)}{m_B^2-(p+q)^2}+
\frac1\pi \int\limits_{s_0}^{\infty}ds \frac{\mbox{Im}F^{(OPE)}(s,q^2)}{s-(p+q)^2}\,,
\label{eq:lcsr0}
\end{equation}
where the residue of the $B$-meson pole term contains the product 
of matrix elements $\langle \pi \!\mid \!\bar{u}\gamma_\lambda b \!\mid \! B\rangle $ and 
$\langle B \!\mid j^\dagger_5 \mid\! 0\rangle$, yielding
the product of the form factor $f_{B\pi}^+(q^2)$ and the decay constant $f_B$. For the latter,
the result obtained from the two-point sum rule discussed in the previous lecture 
can be used. Furthermore, on r.h.s. of (\ref{eq:lcsr0}) we also use
the quark-hadron duality approximation, replacing the integral 
over excited states by  the integral over the spectral density of the OPE 
result with an effective threshold  $s_0$. 
Subtracting the integrals from $s_0 $ to $\infty$ from both sides 
of the above relation and performing the Borel transformation 
we finally obtain the desired LCSR  
for the form factor:
\be
 f_B f_{B\pi}^+(q^2)=\frac{1}{2\pi m_B^2}\int\limits_{m_b^2}^{s_0} ds ~
\mbox{Im}F^{(OPE)}(s,q^2)e^{(m_B^2-s)/M^2}\,.
\label{eq:lcsr}
\ee
The inputs include the $b$-quark mass  $\overline{m}_b$, $\alpha_s$,  
and the set of pion DA's $\varphi_\pi^{(t)}(u)$, t=2,3,4.
The resulting numerical interval for the form factor 
is formed by the uncertainties due to variation of the input and 
of $M^2$ within the interval where one can trust 
OPE and where simultaneously the contribution of excited states 
remains subdominant. A very detailed numerical analysis
of this sum rule can be found in \cite{DKMMO,KMOW}.
The effective threshold can be controlled by the $m^2_B$ calculation
from LCSR. The LCSR for the scalar $B\to \pi$ form factor $f^0_{B\pi}$ 
is obtained employing the second invariant amplitude in the correlation 
function (\ref{eq:lcsrcorr}).

Let me emphasize that the method discussed here  employs a finite 
$b$-quark mass.  At the same time LCSRs  allow for a systematic transition to the infinite   
heavy-quark mass. This limit described in detail, e.g.,
in the reviews \cite{KR98,Braun_rev97}, reproduces the heavy-mass scaling 
of the form factor at large hadronic recoil
\be 
f^+_{B\pi}(q^2=0)\sim 1/m_b^{3/2}\,,
\label{eq:hlim}
\ee
first predicted in \cite{Chernyak}.
Another important feature of LCSRs is that they contain 
both soft end-point  and hard-scattering contributions
to the form factor.
The hard-scattering part is contained in the  $O(\alpha_s)$ contributions to LCSR,
described by the diagrams with perturbative gluon exchanges.
The soft end-point mechanism originates from the part of OPE 
that do not contain gluon exchanges, and is dominated by the leading order diagram.  
It is therefore not surprising that the hard scattering part is suppressed, supporting the dominance of the end-point mechanism for the form factor.
\begin{figure}[t]
\centerline{
\includegraphics[width=0.45\textwidth]{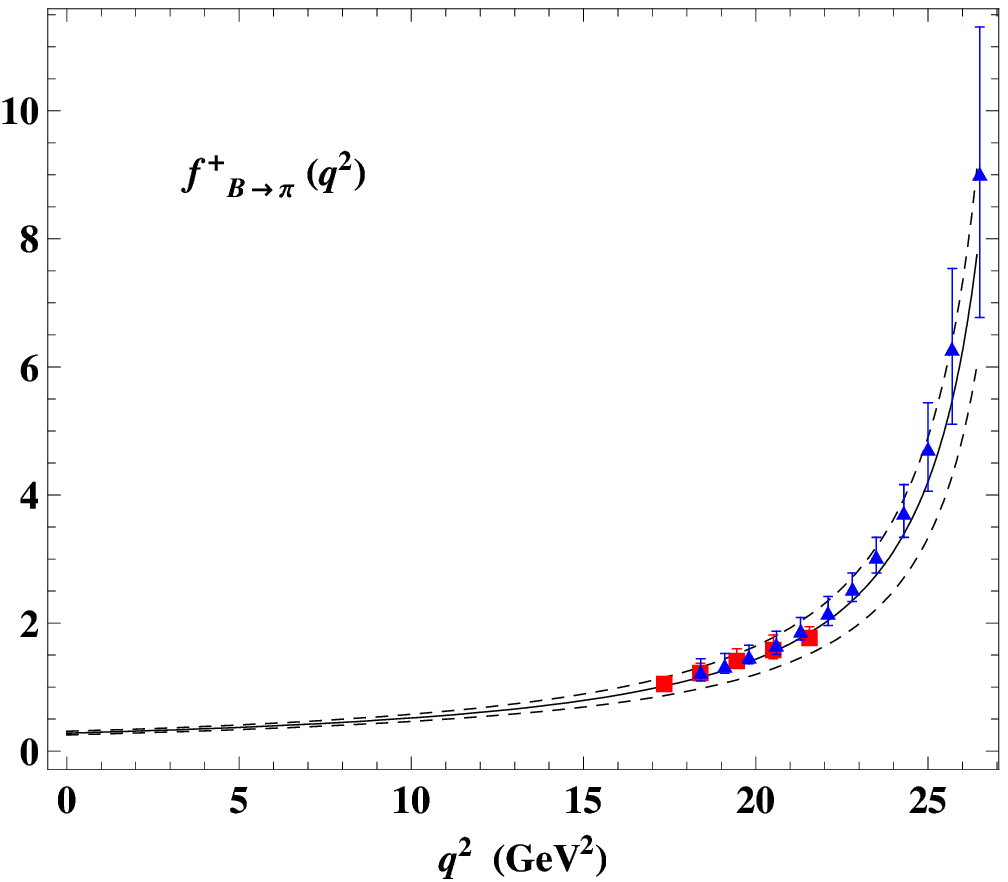}\hspace{1cm}
\includegraphics[width=0.45\textwidth]{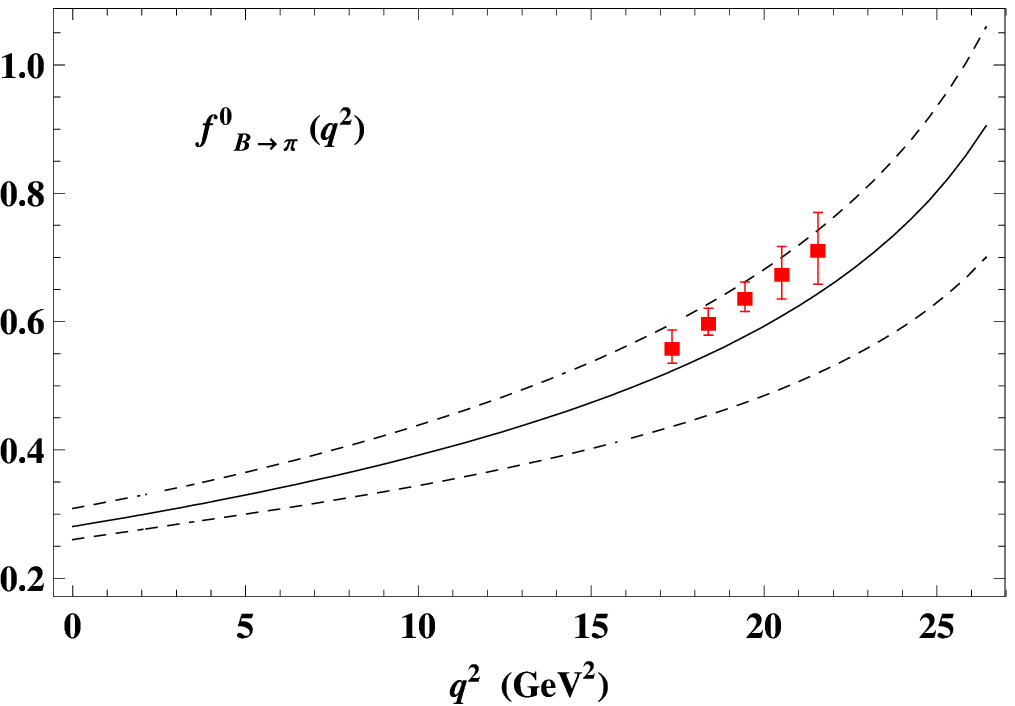}\hspace{1cm}
}
\caption{ \small LCSR results \cite{KMOW} for $B\to \pi$ form factors, extrapolated to 
$q^2>12$ GeV$^2$ in comparison with the lattice QCD predictions \cite{latt_Bpi}}.
\label{fig:fBpires}
\end{figure}

In Fig.~\ref{fig:fBpires} the recent predictions \cite{KMOW}    
of LCSR for both $B\to \pi$ form factors are shown in comparison 
with the lattice QCD results \cite{latt_Bpi}. The sum rules are
used at $q^2<q^2_{max} \simeq 12$ GeV$^2$ and the results are then extrapolated to 
larger momentum transfers with a certain 
analytical parametrization of the form factors \cite{BCL}
as explained in detail in \cite{KMOW}.
Finally, the LCSR results were 
used to evaluate an integral over the weighted form factor squared, 
which, as follows from (\ref{eq:Bpilnu}), is related to the integral
over the partial width: 
 \begin{equation}
\frac{G_F^2}{24\pi^3}\int\limits_0^{q_{max}^2}dq^2p_\pi^3
|f_{B\pi}^+(q^2)|^2= \frac{1}{|V_{ub}|^2\tau_{B^0}}
\int\limits_0^{q_{max}^2} dq^2\frac{d{\cal B}(B\to \pi\ell
\nu_\ell)}{dq^2}\,, 
\label{eq:vub}
\end{equation}
This relation together with the 
measurements of the integrated partial width of $B\to \pi\ell\nu_\ell$ 
were used to extract $|V_{ub}|$. 

Simple replacements $b\to c$ and the adjustment of light quark
flavours in the underlying 
correlation function (\ref{eq:lcsrcorr}) allows to obtain the 
LCSR's for $D\to \pi,K$ form factors \cite{KKMO} employing the same OPE
diagrams. 
In this case, only a narrow region above $q^2= 0$ is accessible with LCSR's.
The $SU(3)_{flavour}$ symmetry violation is encoded in the Gegenbauer moments
of the kaon $a_n^K$ , in particular, the odd moments with $n=1,3,...$ 
have to be added in the  expansion (\ref{eq:Gegenb}). The results for the form factors were used 
in \cite{KKMO} to extract $V_{cs}$ and $V_{cd}$ from the data on $D\to \pi(K) \ell \nu_\ell$  decays.

\subsection{ Alternative sum rules with $B$-meson DA's}

\begin{wrapfigure}{r}{0.35\textwidth}
\centerline{
\includegraphics[width=0.3\textwidth]{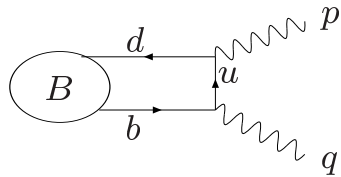}\hspace{1cm}
}
\caption{ \small Correlation function with B-meson DA's 
}
\label{fig:BDAcorr}
\end{wrapfigure}

The positions of the $B$-meson interpolating current and pion in the correlation function 
(\ref{eq:lcsrcorr}) can be exchanged, introducing a new, vacuum-to-$B$ correlation function, 
in which  the $B$ meson is represented by an on-shell state 
and the pion is replaced by an interpolating 
quark current, as shown in Fig.~\ref{fig:BDAcorr}.
Here $q$ is the momentum transfer 
in the weak $b\to u$ transition current and $p$ is the external momentum of the 
light-meson interpolating current, whereas $p_B=p+q$ with $p_B^2=m_B^2$ is the 
$B$-meson momentum.  

This approach was initiated in \cite{KMO}  (see also \cite{FHdF}). Its main advantage is an easy 
extension to other light hadrons, also the non-stable ones. It is
relatively easy to obtain LCSRs for the $B$-meson transition form factors to light 
vector, scalar or axial mesons, by simply varying the quantum numbers of the 
interpolating current and adjusting the quark-hadron duality ansatz.

The description in terms of the light-cone OPE is done 
in the framework of heavy-quark  effective theory (HQET). 
The 4-momentum of $b$ -quark and $B$ meson are represented 
as a sum of the static component and residual momentum: e.g, $p_B=p+q=m_bv+k$ 
where $v$ is the velocity 4-vector. After the transition  to HQET the 
vacuum-to-$B$ correlation function is independent of the scale $m_b$. 
In this effective theory the following  definition \cite{GN,BFeld00}
of the vacuum-$B$  matrix element is used 
\begin{eqnarray}
&&
\langle 0|\bar{q}_{2\alpha}(x)[x,0] h_{v\beta}(0)
|\bar{B}_v\rangle
\nonumber \\ 
&&
= -\frac{if_B m_B}{4}\int\limits _0^\infty 
d\omega e^{-i\omega v\cdot x} 
\left [(1 + \DS v)
\left \{ \phi^B_+(\omega) -
\frac{\phi_+^B(\omega) -\phi_-^B(\omega)}{2 v\cdot x}\DS x \right \}\gamma_5\right]_{\beta\alpha}\,,
\label{eq:BDA}
\end{eqnarray}
where  $h_{v\beta}$ is the effective field, $\alpha,\beta$ are Dirac indices. The functions $\phi_\pm^B(\omega)$ are the $B$-meson two-particle DA's and $\omega$ is the light-quark momentum fraction which formally (in the infinite heavy quark limit) varies up to $\omega=\infty$, however, in all realistic models is limited by $\omega\sim \bar{\Lambda}$ where $\bar{\Lambda}$ is the mass difference
introduced in Eq.~(\ref{eq:barL}). 
More details on $B$-meson DA's can be found in 
the review \cite{Grozin_lect}.
These DA's were used earlier in the context 
of factorization approach to the heavy-light form factors in HQET
\cite{BFeld00}. In addition, the diagram with soft gluon 
emitted from $u$ quark in the correlation function was taken into account, generating the 
three-particle DA's. Their detailed discussion can be found in the 
second paper in \cite{KMO}.  

The rest of LCSR derivation follows the same way as in the case of 
pion DA's. The OPE in terms of $B$-meson DA's 
is matched to the dispersion relation in the variable $p^2$ which is 
the invariant momentum squared of the light-meson interpolating current.
The accuracy of resulting LCSR's for $B\to \pi,K,\rho,K^*$ form factors 
obtained in \cite{KMO}  is still lower than for the 
conventional sum rules. One reason is that 
the key nonperturbative input parameter, the inverse moment:
$\frac{1}{\lambda_B(\mu)}=\int_0^\infty d\omega \frac{\phi_{+}^B(\omega,\mu)}{\omega}$
is not yet accurately determined. 
Two-point QCD sum rules in HQET predict $\lambda_B( 1~ \mbox{GeV})=460\pm 110 ~\mbox{MeV}$ \cite{BIK}. This parameter
is accessible in the photoleptonic'  $B\to \gamma \ell \nu_\ell$ decay 
(for recent  analyses see \cite{BR} and \cite{BK}). 
Another reason is that the radiative gluon corrections to the correlation 
function in Fig.~\ref{fig:BDAcorr} are still missing.
Therefore, the LCSR's with B-meson DA's have a room for improvement.
Finally, let me quote another important application of this method
\cite{FKKM} to  $B\to D^{(*)}$ form factors. The sum rules were obtained 
from the same correlation function as in Fig.~\ref{fig:BDAcorr}
replacing the light quark in the correlation function by a
$c$ quark - another manifestation of the flexibility and universality of the method.

\subsection{Heavy baryon form factors and $\Lambda_b\to p\,\ell\nu_\ell$}

The LCSR method for B-meson form factors
was also extended to the heavy baryon form factors.  In particular, let me 
briefly outline the recent calculation \cite{KKMW} of the $\Lambda_b\to p $ form factors  employing  the following vacuum-to-nucleon correlation function:
\begin{equation}
\Pi_{\mu(5)}(P,q)=i\int d^4z\ e^{iq\cdot z}\langle 0 |T\left\{\eta_{\Lambda_b}(0),
\bar{b}(z)\gamma_\mu (\gamma_5)u(z)\right\}|N(P)\rangle \,.
\label{eq:LbNcorr}
\end{equation}
\begin{wrapfigure}{r}{0.35\textwidth}
\centerline{
\includegraphics[width=0.27\textwidth]{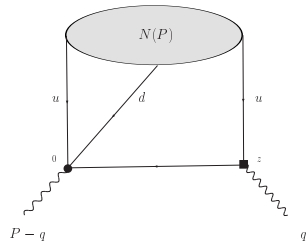}\hspace{1cm}}
\caption{ \small Diagrammatic representation of the correlation function 
with nucleon DA's used to 
derive LCSR's for heavy-to-light baryon
form factors.}
\label{fig:baryoncorr}
\end{wrapfigure}
Here the three-quark heavy-light current operator $\eta_{\Lambda_b}$ 
with quantum numbers of $\Lambda_b$ has a nonvanishing matrix element 
$\langle \Lambda_b\mid\eta_{\Lambda_b}\mid 0\rangle\neq 0$.  It is 
traditionally called the $\Lambda_b$ ``decay constant'', although literally
an annihilation of $\Lambda_b$ would violate the baryon number conservation 
and is absent in SM. Nevertheless, in QCD nothing prevents from introducing the 
auxiliary operator $\eta_{\Lambda_b}$ as an external source of $b$-quark baryonic states. 
As opposed to the meson case, one has a multiple choice 
for constructing the three-quark currents. 
In \cite{KKMW}  two different operators were used:
\begin{equation}
\eta^{({\cal P})}_{\Lambda_b} =
\left(u\,C\,\gamma_5\,d\right)b, ~~ 
\eta^{({\cal A})}_{\Lambda_b} =
\left(u\,C\,\gamma_5\gamma_\lambda\,d\right)\gamma^\lambda\,b \,,
\label{eq:barcorr}
\end{equation}
and the difference 
between the results for the form factors 
was considered as a part of the ``systematic'' uncertainty.

The diagram for the correlation function in LO is shown 
in Fig.~\ref{fig:baryoncorr}, with  the 
on-shell nucleon, carrying the 4-momentum $P$ ($P^2=m_N^2$)
and with the horizontal line denoting the virtual $b$-quark.
The approximation of the free $b$-quark propagation is valid in the kinematical region 
$q^2\ll m_b^2$ , $(P-q)^2\ll m_b^2$, where the integral over $z$ in 
Eq.~(\ref{eq:LbNcorr}) is dominated by small intervals near the light-cone, 
$z^2\sim 0$.

Contracting the virtual $b$-quark fields in 
Eq.~(\ref{eq:LbNcorr}), we recover 
new nonperturbative objects: the nucleon DA's. Their  
definitions and properties were worked out in \cite{BraunetalN}, where
also the LCSR's  for nucleon electromagnetic form factors were obtained. 
The latter sum rules are described by the same diagram of Fig.~\ref{fig:baryoncorr} with a light $u,d$ quark in the horizontal line. The definition of DA's is schematically  given by the following decomposition
of the vacuum-nucleon matrix element:
\begin{eqnarray}
\langle 0 | \epsilon^{ijk} u_\alpha^i(0) u_\beta^j(z
) d_\gamma^k(0) | N(P)\rangle = 
\sum_t \mathcal{S}^{(t)}_{\alpha\beta\gamma} 
\times \int d x_1 d x_2 d x_3
 \delta(1-\sum_{i=1}^{3} x_i)e^{-i x_2 P \cdot z }
F_t(x_i)\,,
\label{eq:buclDA}
\end{eqnarray}
where the expansion goes over twist $t=3,4,5,6$ of light-quark operators
and contains 27 DA's $F_t(x_i)$ depending on 
the shares $x_{1,2,3}$ of the nucleon momentum.

The hadronic dispersion relation for the correlation function (\ref{eq:LbNcorr})
aimed at isolating the ground-state $\Lambda_b$-pole contribution also 
has its peculiarities. The baryonic quark currents  not only interpolate  
the ground states  but also their counterparts with the opposite $P$ parity.
In our case, the $\Lambda_b^*$ baryon 
with $J^P=1/2^-$ located  at $m_{\Lambda^*_b}\simeq m_{\Lambda_b} + (200\div 300)$ MeV,
should also be counted as a ground state 
in the hadronic spectrum.  Therefore we have to include this state in the
resulting dispersion relation separately from the excited states:
\begin{eqnarray}
\Pi_{\mu(5)}(P,q)=  
\frac{\langle 0|\eta_{\Lambda_b}| \Lambda_b\rangle\langle 
\Lambda_b |\bar{b} \gamma_\mu(\gamma_5)u|N\rangle }{m_{\Lambda_b}^2-(P-q)^2} 
+ 
\frac{\langle 0|\eta_{\Lambda_b}| \Lambda^*_b\rangle\langle 
\Lambda^*_b |\bar{b} \gamma_\mu(\gamma_5)u|N\rangle}{m_{\Lambda_b^*}^2-(P-q)^2}
+ \!\!\int\limits_{s_0^h}^{\infty} \frac{ ds\,\rho_{\mu(5)}(s, q^2)}{
s -(P-q)^2 }.
\label{eq:lambdaBdisp}
\end{eqnarray}
In \cite{KKMW} a simple procedure was introduced to eliminate the 
$\Lambda^*$ baryon term in the dispersion relation by forming  
linear combinations of kinematical
structures in the correlation function.
The $\Lambda_b$ term contains the product of decay constant and 
the transition form factors.
There are altogether six form factors of $\Lambda_b\to p $ transition, 
actually their definition is very similar to the familiar one in the 
nucleon $\beta$ decay. The three form factors for the vector part of the weak transition current are defined as:
\begin{eqnarray}
\langle \Lambda_b(P-q)| \bar{b} \,\gamma_\mu\, u | N(P)\rangle
=
\bar{u}_{\Lambda_b} (P-q) \bigg\{f_1(q^2)\,\gamma_\mu
+ i\frac{f_2(q^2)}{m_{\Lambda_b}}\,\sigma_{\mu\nu}q^\nu
+\frac{f_3(q^2)}{m_{\Lambda_b}}\,q_\mu\bigg\}u_N(P) \,,
\label{eq:lambdaB}
\end{eqnarray}
For the axial vector current one has to replace in the above:
$\gamma_\mu \to \gamma_\mu\gamma_5 $ and $f_i(q^2) \to g_i(q^2)$.

The resulting sum rule for each form factor  is obtained 
in a standard way. The result of the diagram calculation 
in terms of nucleon DA's is matched to the dispersion  relation
and the quark-hadron duality approximation in the $\Lambda_b$ channel is employed.   
The decay constant of $\Lambda_b$ is  
estimated from the QCD sum rules for the two-point vacuum correlation functions of the $\eta_{\Lambda_b}$ current and its conjugate.
\begin{figure}[t]
\begin{center}
\includegraphics[width=0.38\textwidth]{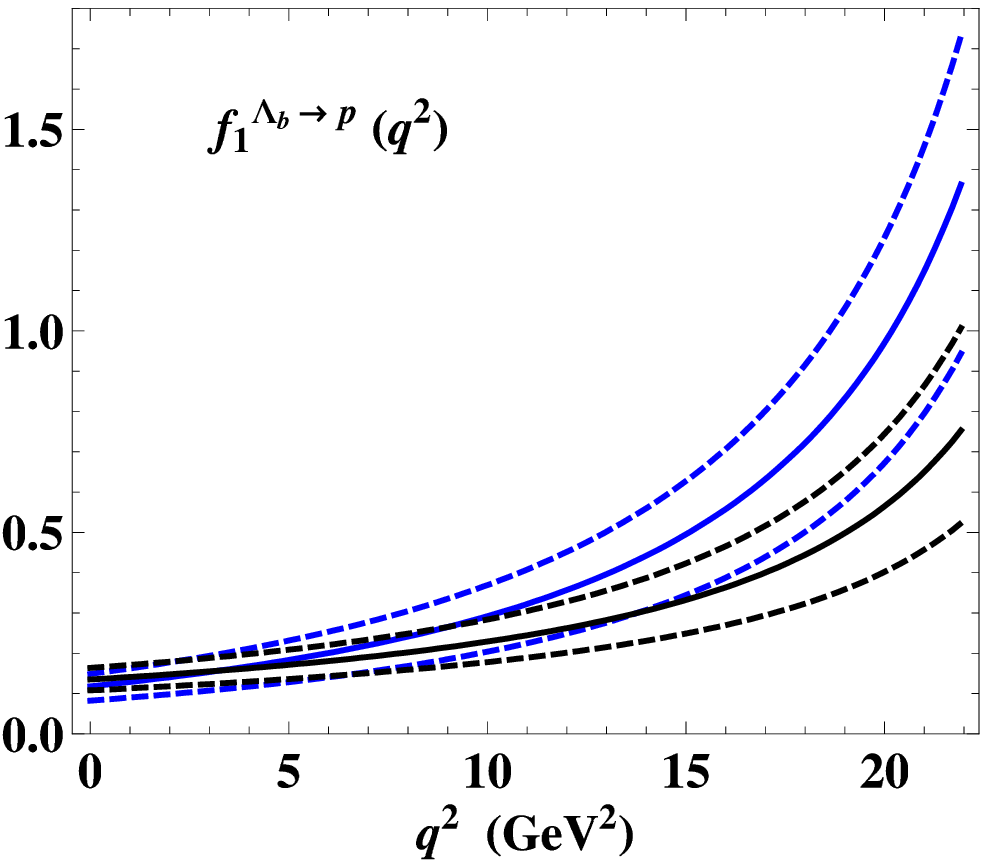}\hspace{1cm}
\includegraphics[width=0.40\textwidth]{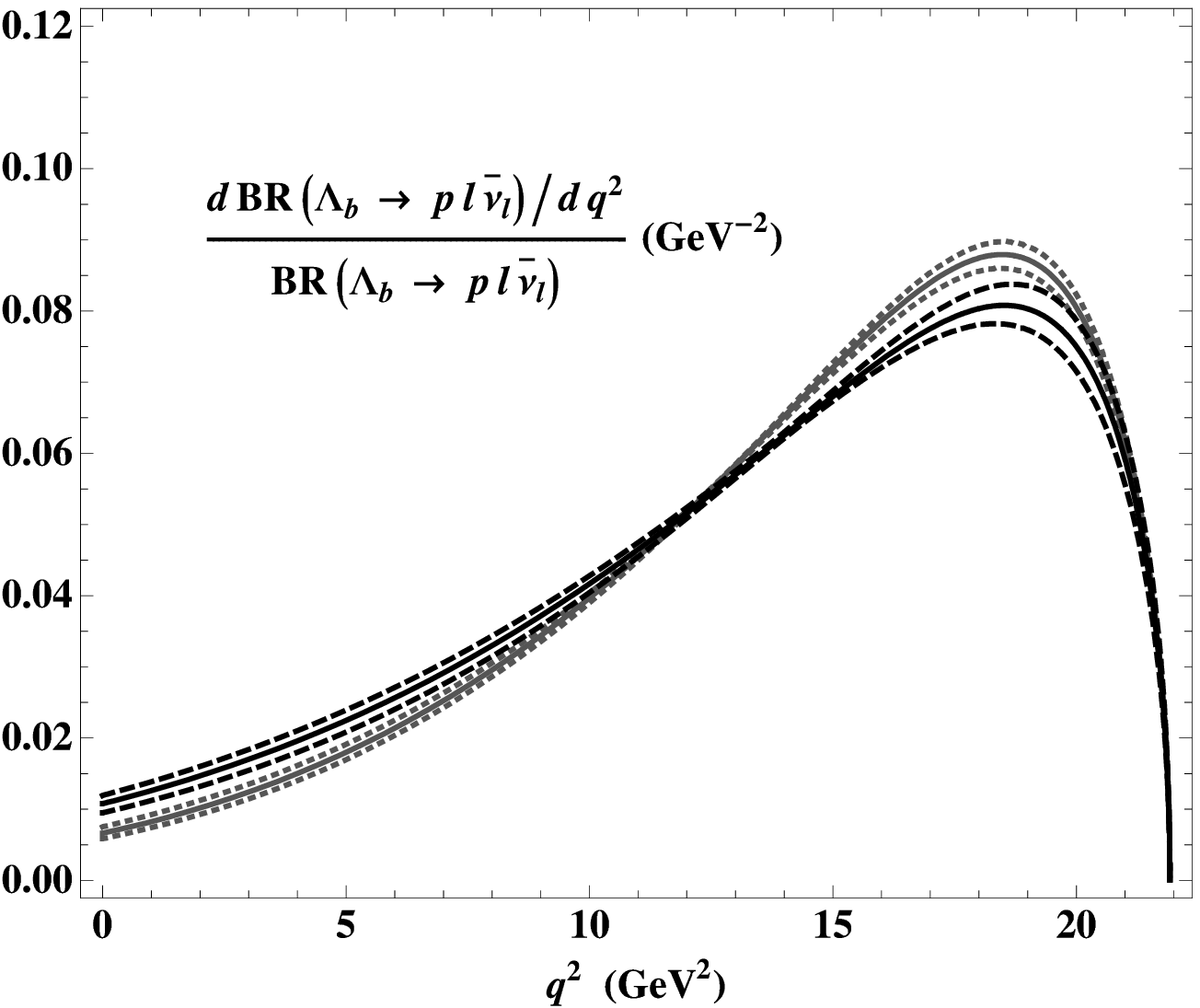}\\
{\small (a) \hspace{5.4cm} (b)}
\end{center}
\caption{\small (a) - one of the $\Lambda_p\to N$ form factors predicted from LCSR
\cite{KKMW}; the spread between solid (solid and dashed) lines
indicates  the difference due to the choice of   
$\Lambda_b$ currents (uncertainties due to the input variation); 
(b) - normalized differential width of $\Lambda_b\to p \ell\nu_\ell$ calculated using LCSR form factors.}
\label{fig:Lbwidth}
\end{figure}

The kinematical region  of the $\Lambda_b\to p \ell\nu_\ell$ 
semileptonic decay,
$ 0\leq q^2 \leq (m_{\Lambda_b}-m_N)^2$, is only 
partly covered by the LCSR calculation. The OPE is not reliable
at large $q^2$, typically at $q^2 > 12-14$ GeV$^2$ because 
the virtual three-quark $bud$-state approaches the 
hadronic threshold in the $q^2$ channel.
The numerical results obtained in \cite{KKMW}
include the form factors at $q^2\leq 11 $ GeV$^2$ calculated with the
universal inputs  including the $b$-quark mass  and a few parameters 
determining the nucleon DA's.
To improve LCSRs one also has to calculate the 
radiative gluon corrections to the correlation function which is however 
technically very challenging.

In Fig.~\ref{fig:Lbwidth} (left) one of the vector form factors is plotted, where 
the analytical parametrization \cite{BCL} fitted to the 
LCSR prediction at low $q^2$ is used to extrapolate this form factor
to the whole region of momentun transfer.   
One observes a reasonable agreement between the sum rules with different 
$\eta_{\Lambda_b}$-currents. 
The $\Lambda_b\to p \ell \nu_\ell$ decay width measurements combined with 
the calculated  form factors
provide an alternative source of $|V_{ub}|$ determination.
\newpage

\section{\!\!\!\!\!\!.~Lecture: Hadronic effects in $B\to K^{(*)}\ell^+\ell^-$}

In this lecture I will discuss
a more complex problem of calculating the hadronic input for exclusive 
flavour-changing neutral current (FCNC) decays. As we shall see, the LCSRs
provide not only the form factors but also nonlocal hadronic
matrix elements specific for these decays.

\subsection{FCNC transitions and nonlocal hadronix matrix elements}
\begin{figure}[b]
\begin{center}
\includegraphics[width=0.55\textwidth]{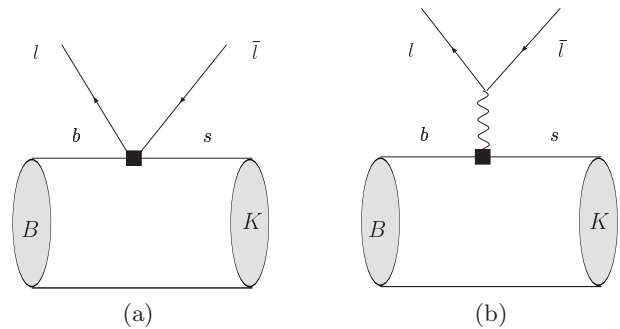}\\
{\small (a) \hspace{4.0cm} (b)}
\end{center}
\caption{\small Hadronic matrix elements of FCNC operators $O_{9,10}$ (a) and $O_7$ (b) 
in the $B\to K \ell^+\ell^-$ decays.}
\label{fig:BKlldiags}
\end{figure}
The $b\to s\, \ell^+\ell^-$ FCNC transitions, observed 
in the form of exclusive decays $B\to K^{(*)}\ell^+\ell^-$,  
are intensively studied at LHC and $B$ factories.
The main interest in these decays is their sensitivity to
the contributions of new heavy particles.
In SM the $b\to  s \ell^+\ell^- $ transitions are described by  
an effective Hamiltonian
\be
H_{eff}= -\frac{4G_{F}}{\sqrt{2}}V_{tb}V_{ts}^{*}
{\sum\limits_{i=1}^{10}} C_{i}(\mu) O_{i}\Big|_{\mu\sim m_b}\,,
\label{eq:Heff}
\ee
where the loop diagrams with heavy SM particle ($t,Z,W$)
are absorbed in the Wilson coefficients $C_i$. The lighter fields, including the $b$ quark field, form effective local operators $O_i$. 
The $B\to K^{(*)}\,\ell^+\ell^-$ decay amplitude 
\ba
 A(B \to K^{(*)} \ell^+\ell^-)=
\frac{G_{F}}{\sqrt{2}}V_{tb}V_{ts}^{*}
\sum\limits_{i=1}^{10} C_{i}(\mu)\, 
 \langle K^{(*)} \ell^+\ell^-\mid O_i\mid B\rangle\Big|_{\mu\sim m_b}
\label{eq:ABKll}
\ea
is written formally as a sum of matrix elements of effective 
operators between  the initial and final states, weighted by their Wilson  coefficients.
The dependence on the scale $\mu$ 
indicates the separation of gluon radiative corrections with momenta larger and smaller
than $\mu$ between the Wilson coefficients and hadronic matrix elements, respectively.

In the above, the dominant contributions to the amplitude (\ref{eq:ABKll})
are given by the operators 
\be
O_{9(10)}=\frac{\alpha_{em}}{4\pi}
[\bar{s}_L\gamma_\mu b_L]\ell\gamma^\mu (\gamma_5)\ell,~~
O_{7\gamma}=-\frac{e m_b}{16\pi^2}[\bar{s}\sigma_{\mu\nu}(1+\gamma_5)b]F^{\mu\nu}
\label{eq:O9}
\ee
with large coefficients $C_9(m_b)\simeq 4.2$, $C_{10}(m_b)\simeq-4.4$
and $C_7(m_b)\simeq -0.3$.
The corresponding diagrams are shown in Fig.~\ref{fig:BKlldiags}. 
The new physics effects can substantially modify the coefficients $C_{9,10,7,...}$,
and/or add new operators with different spin-parity combinations. 
In the contributions of $O_{9,10,7}$, the leptons are factorized out from the 
matrix elements in (\ref{eq:ABKll}) and the only hadronic input  
one needs are the $B\to K^{(*)}$  form factors. The latter can be calculated with LCSR
methods  considered in the previous lecture. 

However, at this stage the problem of 
determining the hadronic input in  $B\to K^{(*)}\,\ell^+\ell^-$ 
is not yet  solved. 
Note that the effective Hamiltonian (\ref{eq:Heff}) also contains effective 
operators without leptons or photon:
the gluon-penguin  $O_{8g}=-\frac{m_b}{8\pi^2}\bar{s}\sigma_{\mu\nu}(1+\gamma_5)bG^{\mu\nu} $,  4-quark penguin operators\,$O_{3-6}$ with small Wilson coefficients  
and, most importantly, 
the current-current operators 
$O^{(c)}_{1}=[\bar{s}_L\gamma_\rho c_L][\bar{c}_L\gamma^\rho b_L]$
and $O^{(c)}_{2}=[\bar{c}_L\gamma_\rho c_L][\bar{s}_L\gamma^\rho b_L] $
of the ``ordinary'' weak interaction, 
with large coefficients $C_1(m_b)\simeq 1.1$ and $C_2(m_b)\simeq -0.25$,
respectively 
\footnote{~The same
operators with $u$ quarks are strongly suppressed by the CKM factor and 
therefore usually neglected in $b\to s$ transitions.}.
These operators also contribute to the $b\to s \ell^+\ell^-$  transition. 
In a combination with weak interaction, 
the lepton pair in the final state is electromagnetically 
emitted from one of the quark lines. The main problem 
is that the average distances between the photon emission
and the weak interaction points are not necessarily short,
hence these additional contributions to the decay amplitude
are essentially nonlocal, and cannot be simply reduced 
to the form factors. 

The following decomposition of the 
decay amplitude in terms of hadronic matrix elements:
\ba
A(B \to K^{(*)} \ell^+\ell^-)=
\frac{G_{F}}{\sqrt{2}}V_{tb}V_{ts}^{*}\frac{\alpha_{em}}{2\pi}
\Bigg[(\bar{\ell}\gamma^\rho\gamma_5\ell\big) C_{10}\,
 \langle K^{(*)}|\bar{s}\gamma_\rho(1-\gamma_5) b|B\rangle 
\nonumber\\
+(\bar{\ell}\gamma^\rho\ell\big)\Big(
C_9\, \langle K^{(*)}|\bar{s}\gamma_\rho b|B\rangle 
+C_7\,\frac{2(m_b+m_s)}{q^2}q^\nu \langle K^{(*)}|\bar{s}i\sigma_{\nu \rho}
(1+\gamma_5) b|B\rangle
\nonumber \\
-\frac{32\pi^2}{q^2}\sum\limits_{i=1,2,...,6,8}C_i
~ {\cal H}^\rho_{i}\Big)\Big]
\ea
includes ``direct'' FCNC contributions proportional to $C_{9,10,7}$ 
multiplied by the  $B\to K^{(*)}$  form factors  and the nonlocal hadronic matrix elements
\be
{\cal H}_{i}^\rho(q,p)=\langle K^{(*)}(p)|i\!\int d^4x\, e^{iqx}\,
T\{ j^\rho_{em}(x), O_i(0)\}|B(p+q)\rangle 
\label{eq:nonloc}
\ee
where $j_{em}^{\rho}=
\sum\limits_{q=u,d,s,c,b} Q_q \bar{q}\gamma^\rho q$ 
is the quark electromagnetic current. The factor $1/q^2$ 
multiplying the nonlocal part of the amplitude is due to the photon propagator 
connecting the quarks  with the lepton e.m. current.

Hereafter,  for simplicity  we consider   the  decay $B\to K \ell^+ \ell^-$ with 
the kaon final state. The  
QCD LCSRs  similar to the ones used to 
calculate $B\to \pi$ form factors
(see the previous lecture),  provide also $B\to K$ form factors.
One has to replace the pion DA's by kaon DA's in the correlation 
function.  Apart from the vector form factor $f_{BK}^+$, the tensor form factor
$f_{BK}^T$ enters due to the $O_7$ operator. 
The LCSR results for all $B\to K$ form factors at $q^2 \leq 12-15 $ GeV$^2$
were updated  in \cite{KMPW}  and the numerical results can be found there.
One obtains values up to 30\% larger than for the corresponding $B\to \pi$ form factors, 
revealing a noticeable violation of $SU(3)_{flavour}$ symmetry.
Our analysis of the $B\to K \ell^+ \ell^-$ amplitude will be constrained 
by the large hadronic recoil region ($q^2< 6-8 $ GeV$^2$)  which is fully covered by LCSR
form factors.
Note that the alternative LCSR's with 
$B$  DA's also provide the $B\to K$ form factors \cite{KMO}, 
albeit with larger uncertainties. 

The $B\to K\ell^+\ell^-$ amplitude,
after inserting the form factors, reads: 
\begin{eqnarray}
A (B \to K \ell^{+} \ell^{-}) = { G_F \over  \sqrt{2} }
{\alpha_{em} \over \pi} V_{tb} V_{ts}^{\ast} \Bigg[
\bar{\ell}\gamma_{\mu} \ell\, p^\mu\bigg( C_9  f^{+}_{BK}(q^2)
\nonumber \\
 + {2 (m_b+m_s) \over m_B+m_K} C_7^{eff} f^{T}_{BK}(q^2) 
+16\pi^2\!\!\!\!\!\sum\limits_{i=1,2,...,6,8}\!\!\!C_i
~{\cal H}^{(BK)}_{i}(q^2)
 \bigg)
+ \bar{\ell} \gamma_{\mu} \gamma_5 \ell \, p^\mu C_{10}
f^{+}_{BK}(q^2) \bigg] \,,
\label{eq:BKllampl}
\end{eqnarray}
where ${\cal H}_{i}^{(BK)}(q^2)$ are the invariant amplitudes 
in the Lorentz-decomposition of (\ref{eq:nonloc}).

\subsection{Anatomy of the nonlocal hadronic matrix elements}

The nonlocal contributions to the decay amplitude (\ref{eq:BKllampl})
can be cast in a form of  corrections to the short-distance 
Wilson coefficient: 
\be
C_9 \to C_9 + \Delta C_9^{(BK)}(q^2), ~~
\mbox{where} ~~\Delta C_9^{(BK)}(q^2)=16\pi^2\!\!\!\sum\limits_{i=1,2,...,6,8}\!\!\!C_i
~\frac{{\cal H}^{(BK)}_{i}(q^2)}{f^+_{BK}(q^2)}\,.
\label{eq:deltaC9def}
\ee
These corrections are $q^2$- and process-dependent
and have to be estimated one by one for separate operators.
The main question we address here is: are the nonlocal matrix elements 
${\cal H}^{(BK)}_{i}(q^2)$  calculable in QCD?

\begin{wrapfigure}{r}{0.35\textwidth}
\centerline{
\includegraphics[width=0.25\textwidth]{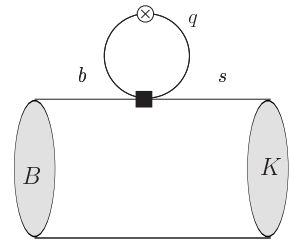}~~~~~
}
\caption{\small The quark-loop diagram of the nonlocal 
contribution to $B\to K\ell^+\ell^-$. The
cross denotes the virtual photon  emission point.}
\label{fig:LOdiags}
\end{wrapfigure}

\begin{figure}[h]
\begin{center}
\includegraphics[width=0.20\textwidth]{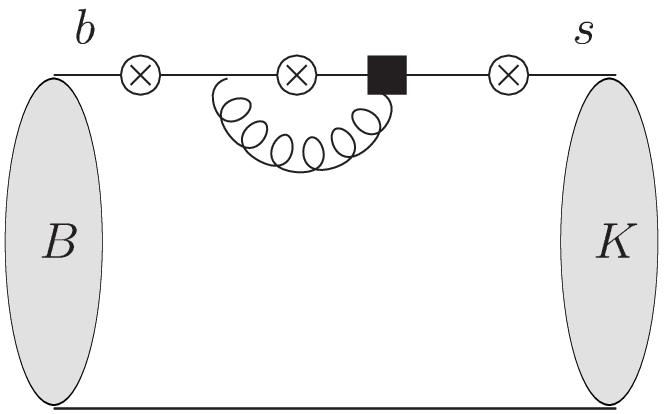}~~
\includegraphics[width=0.20\textwidth]{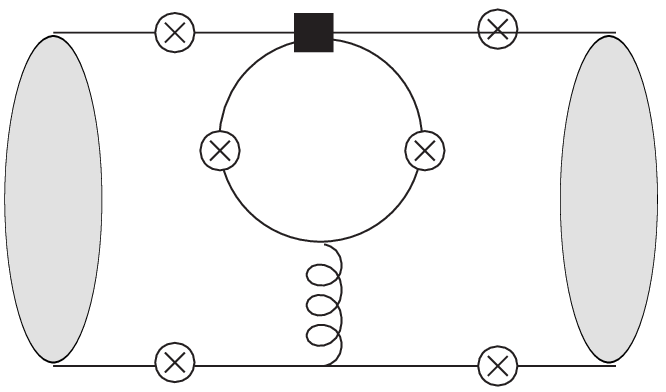}\\
{\small (a) \hspace{4cm} (b)}\\[2mm]
\includegraphics[width=0.20\textwidth]{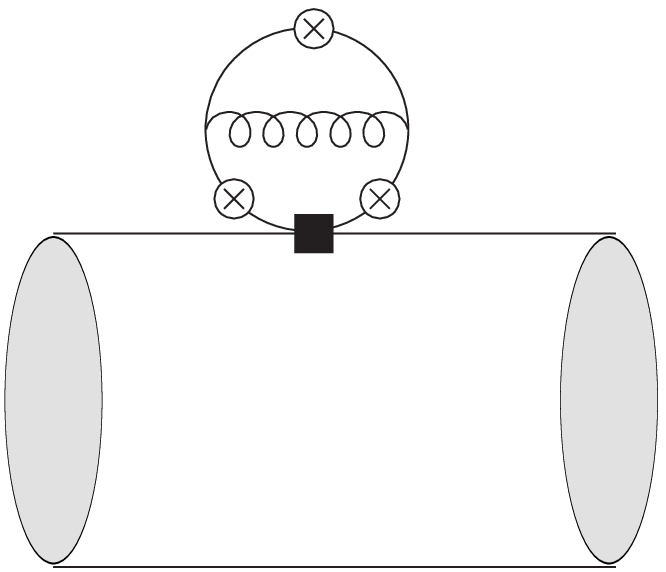}~~
\includegraphics[width=0.25\textwidth]{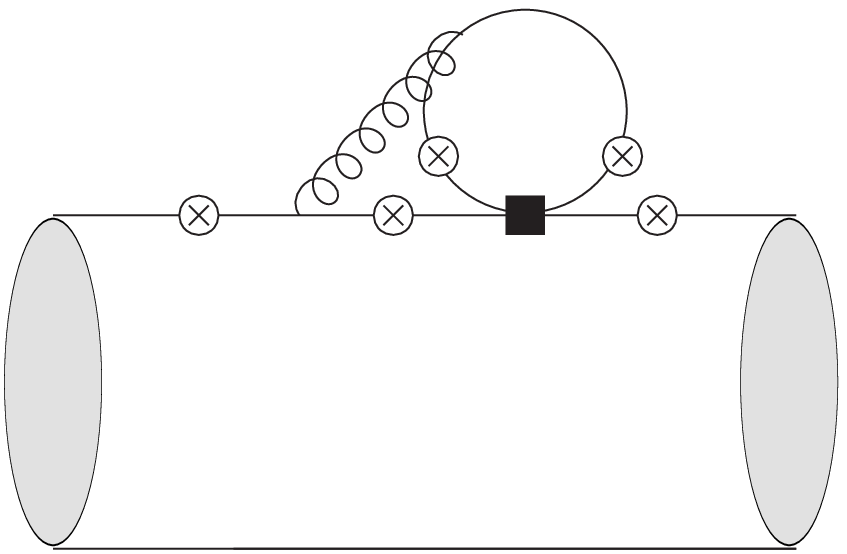}\\
{\small (c)\hspace{4cm} (d)}
\end{center}
\caption{\small Factorizable diagrams with hard gluon exchanges}
\label{fig:NLOdiags1}
\end{figure}

First of all one has to sort out various contributions diagrammatically.
The most important diagram in LO (without additional 
gluons) is in Fig.~\ref{fig:LOdiags}: a virtual photon
emission via intermediate quark loop
originating from the current-current operators $O_{1,2}$ or from quark-penguin operators 
$O_{3-6}$.   In Fig.~\ref{fig:NLOdiags1} the same mechanism is accompanied 
by gluon exchanges including also the gluon penguin contribution.
Not shown is the  mechanism of the weak annihilation 
with virtual photon  emission which has a small impact.  

Calculation of these effects in $B\to K^{(*)} \ell^+\ell^-$ was 
done in the framework of 
HQET and QCD factorization  approach \cite{BFeldS} valid at 
$E_{K^{(*)}}\sim m_b/2$ and  $m_b\to \infty$. 

The results are obtained in the region
of large hadronic recoil (small and intermediate $q^2$). The nonlocal 
amplitudes are expressed in terms of  
$B\to K$  form factors or factorized as a convolution of $B$- and light-meson DA's 
with hard-scattering kernels. 
There are however two problems to clarify. First, at timelike $q^2\sim$ 
a few GeV$^2$,  the virtual photon is emitted via intermediate 
on-shell vector mesons  with the masses 
$m_V=\sqrt{q^2}$  ($ V=\rho,\omega,\phi, J/\psi,...)$ rather than off quarks, hence the accuracy of the 
perturbative treatment has to be assessed. 

\begin{figure}[b]
\begin{center}
\includegraphics[width=0.20\textwidth]{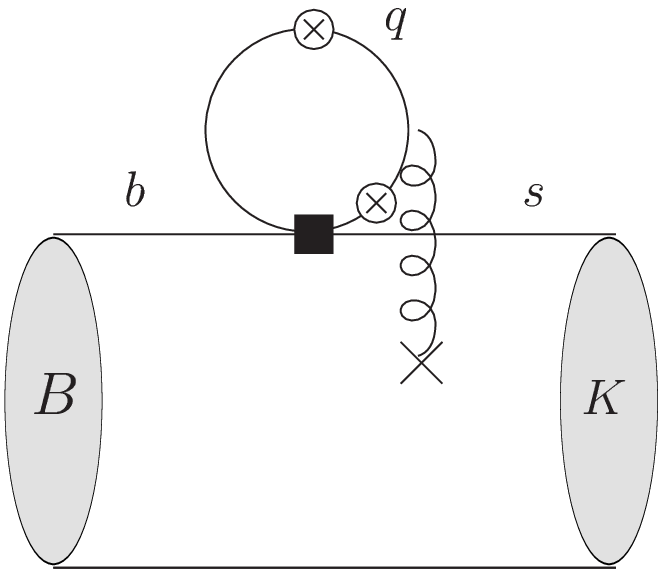}~~
\includegraphics[width=0.20\textwidth]{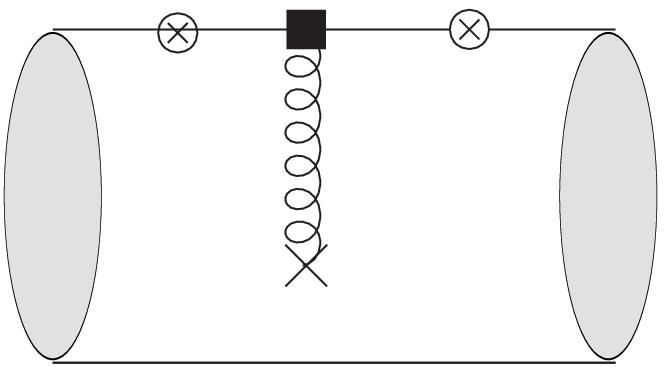}\\
{\small (a) \hspace{3cm} (b)}
\end{center}
\caption{\small Soft-gluon nonfactorizable diagrams}
\label{fig:LOdiags2}
\end{figure}

The second related problem is the role of soft virtual gluons in the
nonlocal amplitudes.
The diagrams shown schematically in Fig.~\ref{fig:LOdiags2}  
are  ``fully nonfactorizable'', i.e.,  with  no possibility to separate a 
hard scattering amplitude from the long-distance one. The whole hadronic 
matix element has to be considered as a nonperturbative  object. 

\subsection{Charm-loop effect and light-cone OPE}

Let me briefly outline the  approach to nonlocal hadronic matrix elements
applied in \cite{KMPW} where the two problems
formulated above were addressed, concentrating on the 
most important (due to large Wilson coefficients) part of the nonlocal 
amplitude generated by the operators $O_{1,2}$.
\begin{figure}[t]
\centerline{
\includegraphics[width=0.65\textwidth]{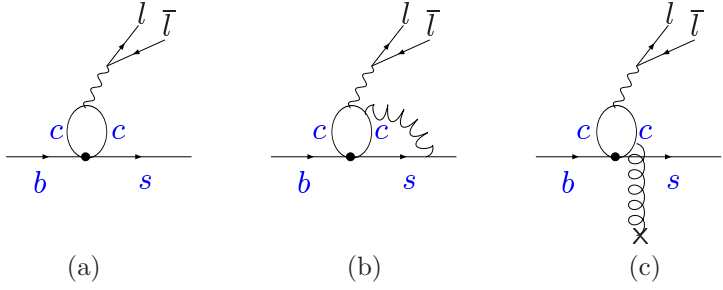}
}
\begin{center}
(a) \hspace{3cm} (b) \hspace{3cm} (c)
\end{center}
\caption{\small $\bar{c}c$-quark loop effect at quark level}
\label{fig:ccdiags}
\end{figure}
This is a combination of the $(\bar{s}c)(\bar c b)$ weak interaction  and 
the $(\bar{c}c)(\bar \ell \ell)$  
e.m.interaction, which effectively leads to $b\to s\ell^+\ell^-$ transition   
due to the fact that the 
charmed quark pair appears in the intermediate state only. 

The leading order diagram shown in Fig.~\ref{fig:ccdiags}(a) 
contains the simple $c$-quark loop  similar to the heavy-light loop 
in the  two-point correlation function considered in the 
first lecture.
Also here the physics depends on the region of the $q^2$ variable. 
At  $q^2 \to m^2_{J/\psi}, ...$   the charm loop turns into 
an on-shell hadronic $J/\psi$ state, and the semileptonic 
decay we are considering becomes a combination
of nonleptonic weak transition $B\to J/\psi K$, followed by 
the e.m. decay $J/\psi\to \ell^+\ell^-$. At larger $q^2$, the
$\psi(2S)$ and other charmonia with $J^P=1^-$, as well  as 
the open-charmed pairs 
contribute, with increasing masses up to the kinematical threshold 
$\sqrt{q^2}=m_B-m_K$.   
To avoid  a ``direct'' charmonium background,
the $q^2$ intervals around $J/\psi$ and $\psi(2S)$ are subtracted 
from the measured lepton-pair mass distributions in $B\to K^{(*)}\ell^+\ell^-$.  
This subtraction does not however exclude the contribution 
of intermediate  virtual ${\bar c}c$ state below the charmonium levels.
Can one use  the  ``loop plus corrections'' ansatz for this contribution
and at which $q^2$?

To investigate  this question, let us  isolate the charm-loop effect 
in the decay amplitude:
\be
A(B\to  K^{(*)} \ell^+\ell^-)^{(O_{1,2})}= -(4\pi\alpha_{em} Q_c)\frac{4G_F}{\sqrt{2}}
V_{tb}V^{*}_{ts}\frac{\bar{\ell}\gamma^\mu \ell }{q^2}
{\cal H}_\mu^{(B\to  K^{(*)})}\,,
\label{eq:cloopampl}
\ee
where the hadronic matrix element:
\begin{eqnarray}
{\cal H}^{(B\to  K)}_\mu(p,q)= i\int d^4xe^{iq\cdot x} \langle  K(p)|
T \Big\{\bar{c}(x)\gamma_\mu c(x)\,,
\Big[ C_1O_1(0)+C_2O_2(0)\Big] \Big\}| B(p+q) \rangle \,,
\label{eq:Amplnonl}
\end{eqnarray}
contains the  $T$-product  of two $\bar{c}c$ operators
\be
{\cal C}^a_{\mu}(q)=\int d^4xe^{iq\cdot x} 
T \Big\{\bar{c}(x)\gamma_\mu c(x) ,\bar{c}_L(0)\Gamma^a c_L(0) \Big\}\,.
\label{eq:cloopexp}
\ee 
As shown in \cite{KMPW}, only at momentum transfers, much lower 
than the charm-anticharm threshold, $q^2\ll 4m_c^2$ one is allowed to 
use the operator-product expansion (OPE), and, importantly  
the expansion is near the light-cone. The dominant region in this $T$-product
is $\langle x^2\rangle \sim 1/(2m_c-\sqrt{q^2})^2$.
In this region the $T$- product of $\bar{c}c$-operators can be expanded  
 near $x^2\sim 0$,  schematically, 
\ba
T\{\bar{c}(x)\gamma_\mu c(x) ,\bar{c}_L(0)\gamma_\rho c_L(0)\} =
C_0^{\mu\rho} (x^2,m_c^2) + \mbox{two-gluon term} + ...
\label{eq:LOterm}
\ea
\ba
T\{\bar{c}(x)\gamma_\mu c(x) ,\bar{c}_L(0)\gamma_\rho\frac{\lambda^a}{2} c_L(0)\} =
\int\limits_0^1 du C_1^{\mu \rho \alpha\beta} (x^2,m_c^2,u) G^a_{\alpha\beta}(ux) + ...
\label{eq:1gluon}
\ea
The leading-order term of this expansion $C_0^{\mu\rho} (x^2,m_c^2)$ 
is reduced to the  simple $\bar{c}c$ loop. Substituting  this term back in the 
decay amplitude (\ref{eq:Amplnonl}),
after  the $x$-integration one obtains
\ba                                                
{\cal O}_\mu(q)=(q_\mu q_\rho- q^2g_{\mu\rho}) 
\frac{9}{32\pi^2}~g(m_c^2,q^2)\bar{s}_L\gamma^\rho b_L \,,
\label{eq:opmu}
\ea
the simple loop function denoted as $g(m_c^2,q^2)$ times the $b\to s$ transition current
(see Fig.~\ref{fig:ccdiags}a).
After taking the hadronic matrix element we recover the {\it factorizable} 
part of the amplitude: 
\ba
 \Big[{\cal H}^{(B\to  K)}_\mu(p,q)\Big]_{fact}= \left(\frac{C_1}{3}+C_2\right)
 \langle  K(p)|{\cal O}_\mu(q)| B(p+q) \rangle \,,
\label{eq:LO}
\ea    
 factorized in the loop function and $B\to K$ form factor
(Fig.~\ref{fig:LOdiags}).
Note that at this level of OPE there is no difference between light-cone 
($x^2\sim 0$) and local ($x\sim 0$) expansion. There are also
perturbative gluon corrections to this operator, one of them
shown in Fig.~\ref{fig:ccdiags}(b). They are factorizable too after taking the 
hadronic matrix elements. For them one can use the results 
 of \cite{BFeldS}, with the only difference that now we  consistently 
avoid the region $q^2\sim 4m_c^2$. 
\begin{wrapfigure}{r}{0.35\textwidth}
\centerline{
\includegraphics[width=0.30\textwidth]{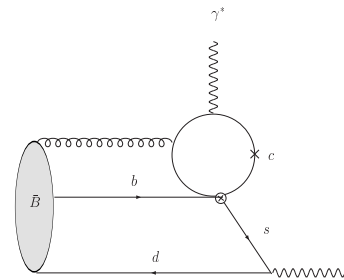}
}
\caption{\small Correlation function used to calculate the nonfactorizable
hadronic matrix element (\ref{eq:hmenonfact}).}
\label{fig:NLOdiags}
\end{wrapfigure}
The genuine nonfactorizable effect is related to the one-gluon term (\ref{eq:1gluon}) 
in the light-cone OPE. It is obtained using the $c$-quark propagator in the 
external gluon field and yields  a new 
{\em nonlocal} operator depicted in Fig.~\ref{fig:ccdiags}(c):
\begin{eqnarray}
\widetilde{{\cal O}}_\mu(q) 
=  \int d\omega\, I_{\mu \rho \alpha\beta}(q,m_c,\omega) 
\bar{s}_L\gamma^\rho 
\delta [ \omega - {(i n_+ {\cal D}) \over 2}]
\widetilde{G}_{\alpha \beta} b_L \,\, ,
\label{eq:nonloc2}
\end{eqnarray}
where the coefficient $I_{\mu \rho \alpha\beta}(q,m_c,\omega) $ 
represents a loop function with gluon insertion
and $n_+$ is the light-like vector defined in $B$ rest frame, 
so that  $q\sim (m_b/2)n_+$. More details can be found in \cite{KMPW}.
The gluon emission  term yields  a new {\it nonfactorizable} hadronic matrix element: 
\ba
 \Big[{\cal H}^{(B\to  K)}_\mu(p,q)\Big]_{nonfact}= 
2 C_1\langle  K(p)|\widetilde{{\cal O}}_\mu(q)| B(p+q) \rangle \,.
\label{eq:hmenonfact}
\ea    
which is not reduced to simple $B\to K$ form factors and corresponds to the diagram in 
Fig.~\ref{fig:LOdiags2}a.  

To calculate the soft-gluon hadronic matrix element (\ref{eq:hmenonfact}),
the method of LCSRs with $B$ meson DA's  outlined in the previous
lecture was used in \cite{KMPW}, introducing a correlation function:
\ba
{\cal F}^{(B\to K)}_{\nu \mu}(p,q)=
i\int d^4y e^{ip\cdot y}
\langle 0| T\{
j_{\nu}^{K}(y) \widetilde{{\cal O}}_\mu(q)\}| B(p+q)\rangle \,.
\label{eq:corrnew}
\ea
The diagram of the correlation function is shown in Fig.~\ref{fig:NLOdiags}
and the OPE contains the 3-particle DAs of $B$ meson.

Summarizing,  the charm-loop effect in $B\to K \ell^+\ell^-$ 
is a sum of two hadronic matrix elements  calculated in QCD,
but this calculation is only  valid at $q^2 \ll  4m_c^2$. 
In \cite{KMPW} the perturbative corrections were not yet included.
Still, to have some idea on the importance of the charm loop effect 
let me quote the value $\Delta C^{(\bar{c}c)}_9(0)= 0.17^{+0.09}_{-0.18}$
obtained for the charm-loop correction to the effective coefficient $C_9$.

\subsection {Hadronic input for  $B\to K \ell^+\ell^-$ decay}
\begin{figure}[t]
\begin{center}
\includegraphics[width=0.5\textwidth]{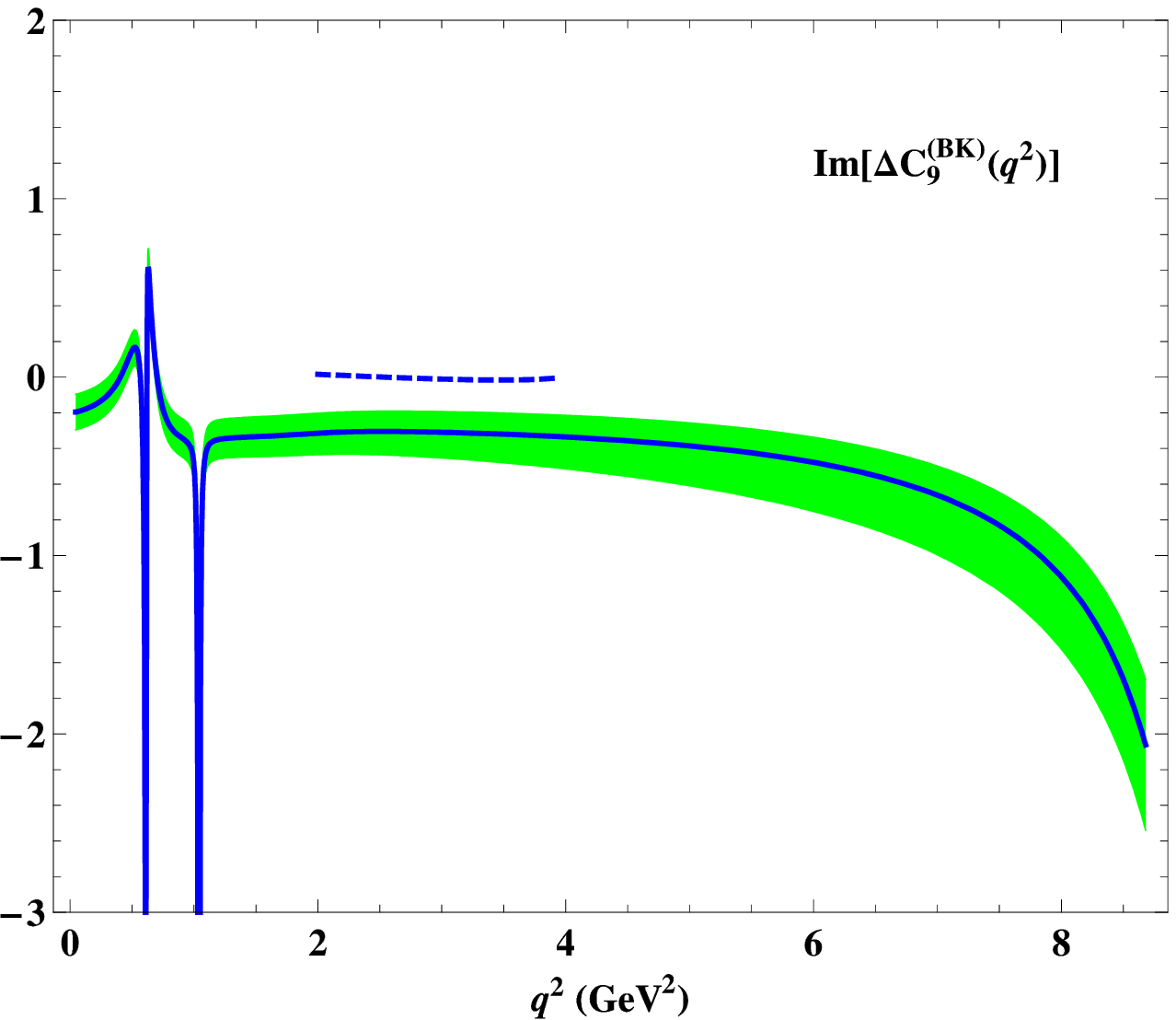}~~
\includegraphics[width=0.5\textwidth]{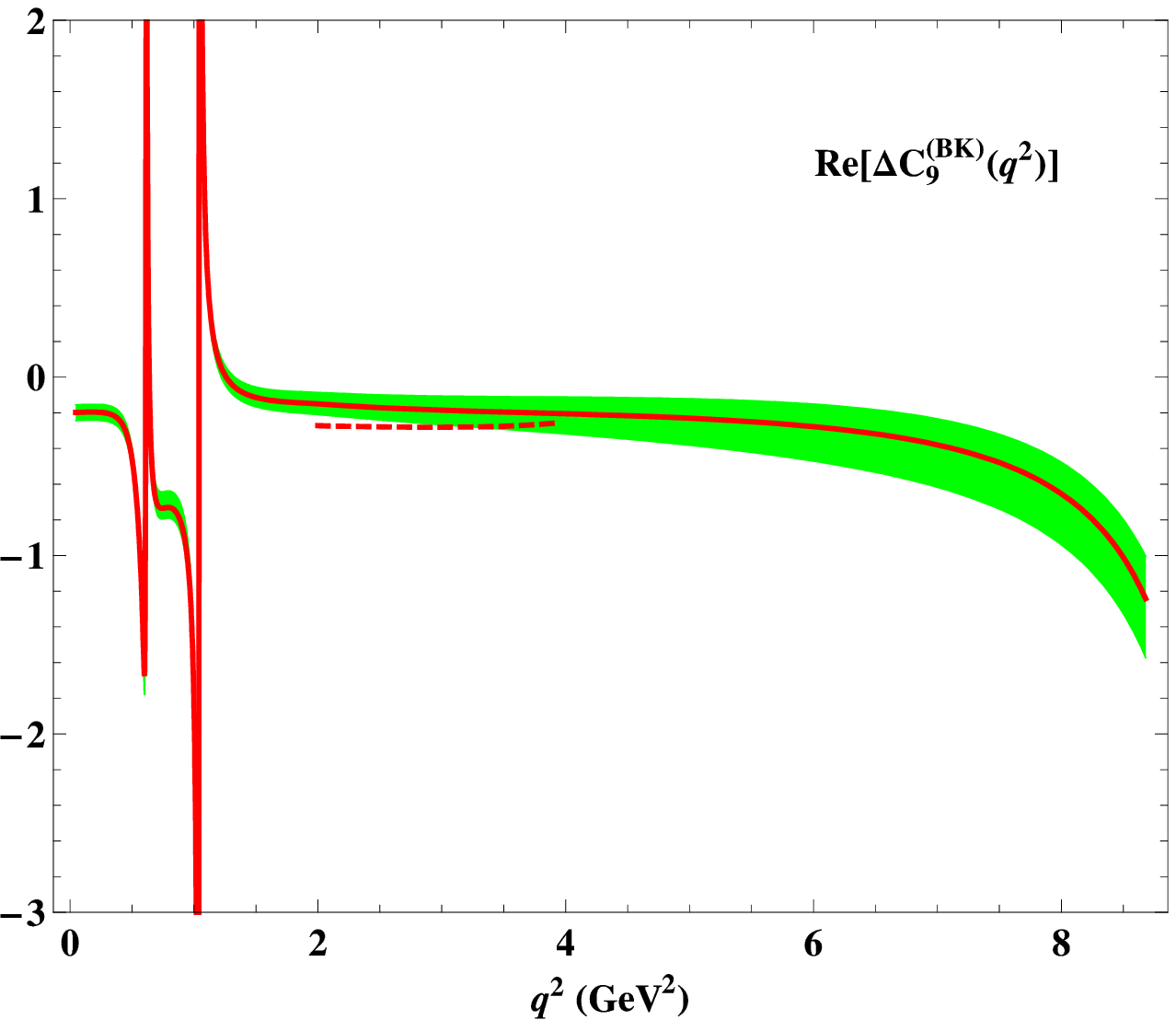}~~\\
{\small (a) \hspace{5.5cm} (b) }
\end{center}
\caption{\small \cite{KMW12} The contribution of nonlocal hadronic effects 
in a form of  correction $\Delta C_9(q^2)$  
to the Wilson coefficient 
$C_9$ in the physical region of $B\to K \ell^+\ell^-$ decay 
obtained from the hadronic dispersion relation, fitted to the QCD calculation
at $q^2<0$. The shaded areas indicate the uncertainty of the predictions. 
The dashed lines are the predictions of QCD factorization \cite{BFeldS}.}
\label{fig:DeltaC}
\end{figure}
Following the method suggested in \cite{KMPW}, in 
\cite{KMW12} a complete ``bookkeeping'' of nonlocal 
contributions to  $B\to K \ell^+\ell^-$ decay amplitude 
was done.  
The soft-gluon effects originating from quark loops 
with various flavours were calculated 
from LCSRs, including also the 
soft-gluon contribution due to the gluon-penguin operator
shown in Fig.~\ref{fig:LOdiags2}b.
In addition also the perturbative gluon exchanges (Fig.~\ref{fig:NLOdiags1}) 
were taken into account employing the results of \cite{BFeldS}.
Note that the latter contributions generate an imaginary part in $\Delta C_9^{(BK)}(q^2)$
as explained in details in \cite{KMW12}.
Furthermore, after  including the photon emission from the light quarks, 
 the $q^2$ region accessible to OPE was shifted 
towards  large negative values of $q^2$, to stay sufficiently far from all hadronic
thresholds.

This calculation was then used for a phenomenological analysis of the 
$B\to K \ell^+\ell^-$ decay.  To access the timelike $q^2$ region 
where OPE is not applicable, the  hadronic dispersion relation in the variable $q^2$ 
was employed \cite{KMPW,KMW12} for the nonlocal hadronic amplitude. 
To illustrate the idea, let us return to the 
previous subsection where only the charm-loop effect was taken 
into account. In this case the dispersion relation contains 
only hadronic states with $\bar{c}c$ flavour content:
\cite{KMPW}:
\ba
{\cal H}^{(B\to K)}(q^2) = {\cal H}^{(B\to K)} (0) + q^2\Big[
\sum_{\psi=J/\psi,\psi(2S)} \frac{f_\psi A_{B \psi K} }{m_\psi^2(m_\psi^2-q^2-
im_\psi\Gamma^{tot}_\psi)}
\nonumber\\
+\int_{4m_D^2}^{\infty} ds \frac{\rho(s)}{s(s-q^2-i\epsilon)}\Big] .
\label{eq:disp}
\ea
The  QCD calculation at small $q^2$
is used to  fit the parameters of this relation and then it  is used
in the timelike region.  In addition,  the absolute values of the residues $|f_\psi A_{B \psi K}|$ 
are fixed from experimental data 
on nonleptonic decays $B\to J/\psi K$,   $B\to \psi(2S) K$ 
and leptonic decays of charmonium \cite{PDG}.

\begin{figure}[t]
\begin{center}
\includegraphics[width=0.5\textwidth]{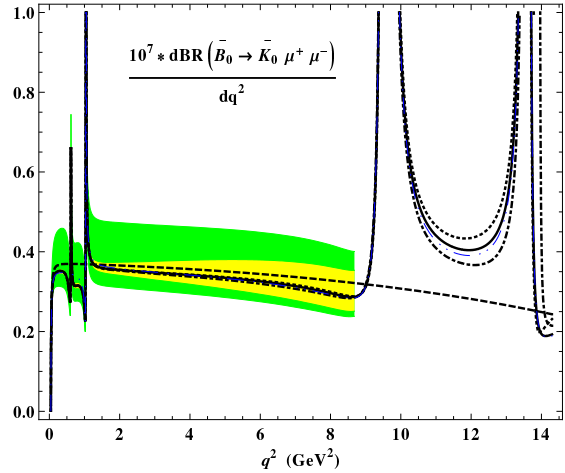}~~\\
\end{center}
\caption{\small  \cite{KMW12} Differential partial branching fraction
of $B\to K \ell^+\ell^-$. The darker (brighter)  shaded area indicates
the uncertainties including(excluding) the one from the $B\to K$ form factors.
The long-dashed line corresponds to the width calculated without nonlocal 
hadronic effects.}
\label{fig:BRBkellell}
\end{figure}

For a full phenomenological analysis of nonlocal amplitude ${\cal H}^{(B\to K)}(q^2)$
in the semileptonic region below charmonium resonances
a   more complete dispersion relation was used in\cite{KMW12},  adding 
vector  mesons with light flavours to the r.h.s. of Eq.~(\ref{eq:disp}).
The main outcome of this analysis is displayed in Fig.~\ref{fig:DeltaC}
where the resulting correction  to $C_9$ due to all nonlocal effects
is plotted, obtained from the dispersion relation fitted to the OPE results
at negative $q^2$ .  Adding these correction to the short-distance 
coefficients and employing the $B\to K$ form factors  from LCSRs
the partial width of $B\to K \ell^+\ell^-$ was predicted in \cite{KMW12}.
It is displayed in Fig.~\ref{fig:BRBkellell}. The influence of nonlocal
effects on the decay observables is very moderate and the form factor uncertainty
still dominates.
For $B\to K^*\ell^+\ell^-$  decay the full analysis still has to 
be done. Hints that the nonlocal hadronic effects in this process
are more pronounced than in the kaon mode come from 
the results for  the charm-loop contribution obtained in \cite{KMPW}. 

Let me emphasize that in future studies of FCNC semileptonic decays 
of $B$ mesons based on more accurate data 
the effects  studied in this lecture are indispensable. Without them
the predictions for SM observables are incomplete.
The methods based on OPE, LCSRs and dispersion relations 
combined with QCD factorization for perturbative contributions
provide  a useful tool to tackle this problem. 

\section*{Acknowledgments}

I am grateful to the organizers of the Helmoltz International Summer School in 
Dubna for an enjoyable scientific event. 
This work is supported  by DFG Research Unit FOR 1873 ``Quark Flavour Physics
and Effective Theories'',  Contract No.~KH 205/2-1.



\begin{footnotesize}

\end{footnotesize}


\end{document}